\documentclass[12pt]{article}

\usepackage{graphicx,color}
\usepackage{amsfonts}
\usepackage{amssymb}
\usepackage{amstext}

\begin{document}



\def\theequation{\thesection.\arabic{equation}}
\def\be{\begin{equation}}
\def\ee{\end{equation}}
\def\ba{\begin{eqnarray}}
\def\ea{\end{eqnarray}}
\def\lb{\label}
\def\nn{\nonumber}

\def\a{\alpha}
\def\b{\beta}
\def\g{\gamma}
\def\d{\delta}
\def\i{\eta}
\def\e{\varepsilon}
\def\l{\lambda}
\def\r{\rho}
\def\s{\sigma}
\def\t{\tau}
\def\o{\omega}
\def\v{\varphi}
\def\x{\xi}
\def\c{\chi}

\def\D{\Delta}
\def\G{\Gamma}
\def\O{\Omega}
\def\L{\Lambda}

\def\Z{\mathbb Z}
\def\Q{\mathbb Q}
\def\R{\mathbb R}
\def\C{\mathbb C}
\def\F{\mathbb F}
\def\H{\mathbb H}

\def\E{{\cal E}}

\def\Rp{\hat{R}(p)}
\def\id{\mbox{\rm 1\hspace{-3.5pt}I}}

\def\p{\hat p}
\def\bo{\mathfrak b}
\def\tp{\tilde p}

\def\eod{\phantom{a}\hfill \rule{2.5mm}{2.5mm}}

\def\bq{\overline{q}}
\def\bU{\overline{U}_q}
\def\bD{\overline {\cal D}}
\def\tU{{\tilde{U}}_q}

\def\uz{\underline z}


\title{\vspace*{-15mm}Zero modes' fusion ring and braid group representations
for the extended chiral \\ ${{su(2)}}$ WZNW model}

\author{
Paolo Furlan$^{a,b~1},\ $
Ludmil Hadjiivanov$^{c,b,d~2},\ $
Ivan Todorov$^{c~3}$}

\date{}

\maketitle

\scriptsize{
$^a$ Dipartimento di Fisica Teorica dell'
Universit\`a degli Studi
di Trieste, Strada Costiera 11, I-34014 Trieste, Italy

$^b$ Istituto Nazionale di Fisica Nucleare (INFN),
Sezione di Trieste, Trieste, Italy

$^c$ Theoretical and Mathematical Physics Division,
Institute for Nuclear Research and Nuclear Energy,
Bulgarian Academy of Sciences,
Tsarigradsko Chaussee 72, BG-1784 Sofia, Bulgaria

$^d$ International School for Advanced Studies
(SISSA/ISAS), via Beirut 2-4, I-34014 Trieste, Italy
}


\begin{abstract}

{A zero modes' Fock space ${\cal F}_q\,$ is constructed for the
extended chiral ${su(2)}\,$ WZNW model. It gives room to a realization
of the fusion ring of representations of the restricted quantum
universal enveloping algebra $\bU=\bU sl(2)\,$ at an even
root of unity, $q^h=-1\,,$ and of its infinite dimensional extension $\tU\,$ by
the Lusztig operators $E^{(h)},\, F^{(h)}\,.$
We provide a streamlined derivation of the characteristic equation for the Casimir invariant
from the defining relations of $\bU\,.$
A central result is the characterization of the Grothendieck ring of both $\bU\,$ and $\tU\,$
in Theorem 3.1. The properties of the $\tU\,$ fusion ring in ${\cal F}_q\,$ are related to the braiding
properties of correlation functions of primary fields of the conformal
$\widehat{su}(2)_{h-2}\,$ current algebra model.}

\end{abstract}

\vspace{0.6cm}



\vfill

\footnoterule
\footnotesize{
$^1$ e-mail address: furlan@trieste.infn.it

$^2$ e-mail address: lhadji@inrne.bas.bg

$^3$ e-mail address: todorov@inrne.bas.bg}

\newpage


\tableofcontents

\section{Introduction}
\setcounter{equation}{0}
\renewcommand\theequation{\thesection.\arabic{equation}}

The {\it extended} $\widehat{su}(n)_k$ {\it chiral
Wess-Zumino-Novikov-Witten} (WZNW) {\it model} can be
characterized as a non-unitary Conformal Field Theory (CFT) which
involves primary states of arbitrary $su(n)\,$ weights $\Lambda\,$
{\it not} restricted to the integrable ones (for which $(\Lambda |
\theta)\le k$ where $\theta\,$ is the highest root) for a positive
integer level $k\,.$
It has been argued at an early stage that the quantum group counterpart
of an integer level $\widehat{su}(n)_k$ WZNW model is the restricted
{\em quantum universal enveloping algebra} (QUEA) $\bU s\ell(n)\,$ at $q\,$ an even
root of unity that is factored by the relations
\be
E_\a^h = 0 = F_\a^h\,,\qquad K_i^{2h} = \id\qquad{\rm for}\qquad h=k+n\qquad (q^h = -1)
\lb{gen-res}
\ee
(see \cite{FK}, Chapter 4; after intermediate sporadic applications, see e.g.
\cite{FHT4}, it was studied more systematically in \cite{FGST1, EGST}). It is a
finite dimensional QUEA that has a finite number of irreducible representations but a
rather complicated tensor product decomposition, partly characterized by its
{\em Grothendieck ring} (GR) which "forgets" the indecomposable structure
of the resulting representations (see Section 2.3 below for a precise definition, and Section 3.4 for a
description of the GR in the present context for $n=2\,$). The interest in the GR structure of
$\bU \equiv \bU s\ell(2)\,$ has been justified by
its relation to the fusion ring of the corresponding CFT model. This relation, noticed by a number of
physicists at the outbreak of interest in quantum groups in the late 1980's was made precise by the
{\em Kazhdan-Lusztig correspondence} (of the 1990's) verified for the logarithmic $c_{1p}\,$ Virasoro model in
\cite{FGST1, FGST2, FHST, Fuchs} and discussed for logarithmic extensions of minimal and $\widehat{s\ell}(2)_k\,$
conformal theories (\cite{FGST3, FGST4, S, S2}). The present paper considers instead the infinite dimensional
{\em Lusztig QUEA} $\tU\,$ that includes the operators $E^{(h)}\,$ and $F^{(h)}\,,$ whose definition is
recalled in Section 2.1, as the true counterpart of the extended chiral WZNW theory.

Our starting point is
the algebra of the zero modes $a=(a^i_\a)\,$
of a chiral group valued field \cite{AF, G, HPT, FHT3} and its Fock space representation. The
{\em quantum matrix} $\,a\,$ intertwines chiral vertex operators (with diagonal monodromy) and quantum group covariant
chiral fields. The resulting {\em quantum matrix algebra} ${\cal A}_q\,$ was studied in the general $U_qs\ell(n)\,$
theories in \cite{FHIOPT} and \cite{Goslar}. In the $U_qs\ell(2)\,$ case, to which the present paper is devoted,
${\cal A}_q\,$ is essentially equivalent to the "twisted oscillator algebra" introduced long ago by Pusz and Woronowicz \cite{PW}.
It is generated by six elements, $a^i_\a \ (i,\a=1,2)\,$ and $q^{\pm\hat p}\,,$ satisfying
$R$-matrix type exchange relations recalled in Section 2.1. The {\em monodromy subalgebra}, introduced in Section 2.2, can be
identified with the commutant of $q^{\pm\hat p}\,$ in ${\cal A}_q\,.$ It gives rise to
the {\em quantum double} that provides, in particular, an extension of $U_qs\ell(2)\,.$ For $q^h=-1\,$ the relations (\ref{gen-res})
(for $n=2\,$) are automatically satisfied in the Fock space ${\cal F}_q\,$ of ${\cal A}_q\,$
(with an $U_qs\ell (2)$ invariant vacuum vector
annihilated by $a^2_\a\,$); more precisely,
\be
{\cal R}_h\, {\cal F}_q = 0\qquad{\rm for}\qquad
{\cal R}_h = \{\, E^h\,,\, F^h\,,\, q^{hH} -q^{-hH}\,,\,q^{hH} + q^{h \hat p}\,\}\ .
\lb{extA}
\ee
Thus, ${\cal F}_q\,$ can be viewed as an $\bU\,$ module. Only a quotient algebra $U_q^F\,$ of $\bU\,$
(with an $(h+1)$-dimensional semisimple centre) is represented faithfully, however.
We shall argue in the present paper that there is a
duality between the (irreducible and) indecomposable Fock space representations of
the infinite (Lusztig) extension $\tU\,$ of $\bU\,$
and the braiding properties of $\widehat{su}(2)_{h-2}\,$ primary fields.
(Clearly, as we view $\tU\,$ as an operator algebra in ${\cal F}_q\,,$
it actually appears as an extension of $U_q^F\,.$)

The monodromy subalgebra of ${\overline{\cal A}}_q = {\cal A}_q / {\cal R}_h\,$ can be identified with the
{\em double cover} $\bD\,$ of $\bU\,$  (Section 2.2). The algebra ${\cal A}_q\,$ (as well as its extension
${\tilde{\cal A}}_q\,$ including $\tU\,$) possesses a
series of nested ideals ${\cal I}_{h} \supset {\cal I}_{2h}\supset\dots\,,$
generated by multiple of $h\,$ powers of $a^i_\a\,$ (Section 2.3). Unlike earlier work \cite{FHT3, Goslar},
here we do not set to zero the maximal ideal ${\cal I}_{h}\,,$ thus admitting indecomposable representations of
$\bU\,$ in the Fock space representation of ${\overline{\cal A}}_q\,$ displayed in Section 2.3.

To make the paper self-contained we review and further elaborate, in Section 3, results of \cite{FGST1, FGST2}
on the {\em Drinfeld map} and the centre ${\cal Z}_q\,$ of $\bU\,,$ and on the realization
(\ref{DR-gen}) of the Drinfeld image ${\mathfrak D}_{2h}\subset {\cal Z}_q\,$ of canonical irreducible characters.
We express the central element $q^{hH}\,$ of $\bU\,$ as a
Chebyshev polynomial of the first kind of the Casimir operator $C\,,$ Eq. (\ref{qhHTh1});
this allows us to derive in a straightforward manner the characterictic equation $P_{2h}(C)=0\,$
from the defining relations of $\bU\,.$
The structure of the fusion ring assumes a particularly simple form when written in terms of the
homogeneous in $a^1_\a\,$ subspaces ${\cal V}_p\,$ of the Fock space
that are indecomposable $\bU$ modules for $p>h\,$ (see Theorem 3.1 in Section 3.4 where we also
characterize the GR of $\tU$). We prove the statement made in \cite{FGST1} that the quotient
of ${\mathfrak D}_{2h}\,$ with respect to the annihilator of its radical
is isomorphic to the fusion ring of the unitary $\widehat{su}(2)_{h-2}\,$ model (Proposition 3.4).

In Section 4 we display the duality between the structure of the indecomposable $\tU\,$ modules
${\cal V}_p\,$ and that of the braid group modules ${\cal S}_4(p)\,$ of $4$-point block
solutions of the Knizhnik-Zamolodchikov equation (of $su(2)\,$ weight $2I_p = p-1\,$). This involves
arrow reversal in the short exact sequences describing the indecomposable structure of dual
representations. A systematic study of what should replace the "Kazhdan-Lusztig correspondence"
(of \cite{FGST1, FGST2}) between the representation categories of $\tU\,$ and of the ${\widehat{su}}(2)_{h-2}\,$
current algebra is left for future work.

\medskip

\section{Chiral $\widehat{su}(2)\,$ zero modes and their Fock space}
\setcounter{equation}{0}
\renewcommand\theequation{\thesection.\arabic{equation}}

We first define, in Section 2.1, $a^i_\a\,$ as $U_q$-covariant ($q$-deformed) "creation and annihilation
operators" and then display, in Section 2.2, their relation to the monodromy of a chiral WZNW field.

\medskip

\subsection{The quantum matrix algebra for $q^h=-1$}

We shall be dealing with the {\em quantum universal enveloping algebra} (QUEA) $U_q\equiv U_q s\ell(2)\,$
defined as a Hopf algebra with generators $E\,,\, F\,$ and $q^{\pm H}\,$ satisfying\footnote{This algebraic structure
was first introduced in $1981$ by P. Kulish and N. Reshetikhin in the context of the $XXZ\,$ spin chain. For a
historical survey and references to original work see \cite{Faddeev}.}
\ba
&&q^H E q^{-H} = q^2 E\,, \qquad q^H F q^{-H} = q^{-2} F\,,\qquad q^{\pm H} q^{\mp H} = \id\,,\nn\\
&&[E , F ] = [H] := \frac{q^H-q^{-H}}{q-q^{-1}}\,,\qquad\qquad q\in {\C}\backslash \{0, \pm 1\}\ ,
\lb{Uqsl2-alg}
\ea
with coproduct $\Delta\,:\, U_q\ \rightarrow\ U_q\otimes U_q\,,$ an algebra homomorphism given on the generators by
\be
\Delta (E) = E \otimes q^H+ \id \otimes E\,,\qquad\Delta (F) = F \otimes \id  + q^{-H}\otimes F  \,,\qquad
\Delta (q^{\pm H}) = q^{\pm H} \otimes q^{\pm H}
\lb{coprUq}
\ee
and with a counit $\e\,: U_q\ \rightarrow\ {\C}\,$ and an antipode (a linear antihomomorphism
$S\,: U_q\ \rightarrow\ U_q\,$) such that
\ba
&&\varepsilon (E)  = 0 = \varepsilon (F)\,,\qquad\, \varepsilon (q^{\pm H} ) = 1\,,\lb{couUq}\\
&&S(E) = - E q^{-H} \,,\qquad\,  S(F) = - q^H F \,, \qquad S(q^{\pm H}) = q^{\mp H}\,.
\lb{antUq}
\ea

We are introducing a deformation ${\cal A}_q$ of
Schwinger's (1952) $SU(2)\,$ oscillator algebra
\cite{Schwinger}
in which the $SU(2)\,$ covariance condition is substituted by $U_q\,$
covariance of $a^i_\alpha\,,\ i,\a=1,2\,,$ expressed by the relation
\be
Ad_X (a^i_\a ) = a^i_\b \,(X^f)^\b_\a\qquad \forall X\in U_q\,.
\lb{AdX}
\ee
Here the superscript $^f\,$ stands for the fundamental (two-dimensional) representation of $U_q\,,$
\ba
&&E^f = \left( \matrix{0&1\cr0&0} \right)\,,\quad
F^f = \left( \matrix{0&0\cr1&0} \right)\,,\quad
\left( q^{\pm H}\right)^f = \left( \matrix{q^{\pm 1}&0\cr0&q^{\mp 1}} \right)\ ,\nn\\
&&[H^f] = H^f = \, \left(\matrix{1&0\cr 0&-1}\right)\ ,
\lb{Xf}
\ea
while the adjoint action $Ad_X\,$ of $X\,$ on $z\in {\cal A}_q\,$ is defined by
\be
Ad_X (z) := \sum_{(X)} X_1\, z\, S(X_2) \qquad
{\rm for}\quad
\Delta (X) = \sum_{(X)} X_1 \otimes X_2\,.
\lb{def-Ad}
\ee
In other words,
\ba
&&q^H a^i_1 = q\, a^i_1 q^H\,,\qquad\qquad\, q^H a^i_2 = q^{-1} a^i_2\, q^H\,,\nn\\
&&[ E , a^i_1 ] = 0\,,\qquad\qquad\qquad   [ E , a^i_2 ] = a^i_1 q^H\,,\nn\\
&&F a^i_1 - q^{-1} a^i_1 F = a^i_2\,,\qquad
F a^i_2 - q\, a^i_2 F = 0\,,\qquad i = 1, 2\ .
\lb{Uqa21}
\ea

The $U_q\,$ {\em quantum matrix algebra} ${\cal A}_q\,$ consistent with these covariance conditions
is defined as an associative algebra with $6\,$ generators, $a^i_\a\,$ and $q^{\pm {\hat p}}\,,$ satisfying
\ba
&&q^{\hat p} a^1_\a = a^1_\a q^{{\hat p}+1}\,,\qquad q^{\hat p} a^2_\a = a^2_\a q^{{\hat p}-1}\,,
\qquad q^{\pm \hat p} q^{\mp \hat p} = \id\,,\lb{ap2}\\
&&a^2_\a a^1_\b = a^1_\a a^2_\b + [{\hat p}]\, {\cal E}_{\a\b}\,,\qquad
a^i_\a a^i_\b\, {\cal E}^{\a\b} = 0\,,\quad i=1,2\,,\nn\\
&&a^2_\a a^1_\b\, {\cal E}^{\a\b} = [{\hat p}+1]\,,\qquad\quad\ \
a^1_\a a^2_\b\, {\cal E}^{\a\b} = - [{\hat p}-1]\,,
\lb{aex2}
\ea
where the $q$-deformed Levi-Civita tensor ${\cal E}\,$ is given by
\be
( {\cal E}_{\a\b} ) =   \left(\matrix{0&-q^{\frac{1}{2}} \cr q^{-\frac{1}{2}}&0}\right) = ({\cal E}^{\a\b})
\qquad ({\rm so\ that}\ {\cal E}_{12} = -q\, {\cal E}_{21}\,)\, ;
\lb{eps1}
\ee
as a result,
\be
{\cal E}^{\a\s}{\cal E}_{\s\b} = - \d^\a_\b\,,\qquad {\cal E}^{\a\s}{\cal E}_{\b\s} =
\left( q^{\tau_3}\right)^\a_\b\,,\qquad
\tau_3 = \left(\matrix{1&0\cr 0&-1} \right)\, .
\lb{eps2}
\ee

The operators $q^{\pm{\hat p}}\,$ commute with $U_q\,,$ which implies
that they are also $Ad$-invariant, i.e. $\, Ad_X (q^{\pm \hat p}) = \e(X)\, q^{\pm \hat p}\quad \forall X\in U_q\,.$
As we shall recall in Section 2.3 below, the eigenvalues of $\hat p\,$ play the role of dimensions of the Fock space
representations of $U_q\,.$ Note that the $U_q\,$ transformation properties (\ref{Uqa21}) of
$a^i_\a\,$ are independent of the upper index $i\,$ while, in accord with (\ref{ap2}),
$\hat p\, a^1_a = a^1_a (\hat p + 1)\,,\ \hat p\, a^2_a = a^2_a (\hat p - 1)\,$ so that
$a^1_\a\,$ and $a^2_\a\,$ play the role of creation and annihilation operators, respectively.

We shall equip $U_q\,$ with a substitute quasitriangular structure\footnote{
A Hopf algebra ${\mathfrak A}\,$ is said to be
{\em almost cocommutative} (see e.g. \cite{CP}), if there exists
an invertible element ${\cal R}\in{\mathfrak A}\otimes {\mathfrak A}\,$
which relates the coproduct $\Delta (x) = \sum_{(x)} x_1\otimes x_2\,$ to its opposite:
${\Delta}^{opp} (x) := \sum_{(x)} x_2\otimes x_1 = {\cal R}\, \Delta (x)\, {\cal R}^{-1}\,.$
An almost cocommutative ${\mathfrak A}\,$ ($\equiv ({\mathfrak A}\,, {\cal R})$)
is {\em quasitriangular} if ${\cal R}\,$ satisfies, in addition,
$(\Delta \otimes id) {\cal R} = {\cal R}_{13} {\cal R}_{23}\,,$
$(id \otimes \Delta ) {\cal R} = {\cal R}_{13} {\cal R}_{12}\,$
(note that the last equations fix the normalization of ${\cal R}\,$).
Quasitriangularity implies the Yang-Baxter equation,
${\cal R}_{12} {\cal R}_{13} {\cal R}_{23} = {\cal R}_{23} {\cal R}_{13} {\cal R}_{12}\,,$
as well as the relations
$(\e\otimes id) {\cal R} = \id = (id\otimes \e) {\cal R}\,,\
(S\otimes id) {\cal R} = {\cal R}^{-1} = (id\otimes S^{-1}) {\cal R}\,,\
(S\otimes S) {\cal R} = {\cal R}\,.$
If $({\mathfrak A}\,, {\cal R})$ is quasitriangular, so is $({\mathfrak A}\,, {\cal R}_{21}^{-1})\,.$}
by introducing the series
\be
{\cal R}_q =  \sum_{\nu = 0}^{\infty} \frac{ q^{-\frac{\nu (\nu -1)}{2}}(-\l)^\nu}{[\nu ]!}\,
F^\nu\otimes E^\nu\, q^{- \frac{1}{2} H\otimes H} \,,\qquad\l := q-q^{-1}\ .
\lb{RUq}
\ee
which plays the role of a universal $R$-matrix but does not belong to $U_q \otimes U_q\,$
(note that Eq. (\ref{RUq}) involves a choice, since
\be
({\cal R}_q)_{21}^{-1} = q^{\frac{1}{2} H\otimes H} \sum_{\nu = 0}^{\infty}
\frac{ q^{\frac{\nu (\nu -1)}{2}}\l^\nu}{[\nu ]!}\,
E^\nu\otimes F^\nu
\lb{RUq21-1}
\ee
has the same properties as ${\cal R}_q\,$ without being equal to it). One can,
following Drinfeld, see e.g. \cite{CP}, give meaning to (\ref{RUq}) and
(\ref{RUq21-1}) by replacing $U_q\,$ with an algebra of formal
power series in $E,F,H$ and $\log\, q\,$ (which would allow, in
particular, to define $q^{\pm\frac{1}{2} H \otimes H}$) and using
an appropriate completion of the tensor product $U_q\otimes U_q\,.$
It is possible, for our purposes, to stay within the
purely algebraic setting (and speak instead of a "substitute
$R$-matrix"). Indeed, in any representation in which either $E\,$ or $F\,$ is nilpotent
and the spectrum of $q^H$ only contains integer powers of $q\,$
(in particular, in all finite dimensional irreducible representations of $U_q\,$
that have a "classical", $q=1$ counterpart), ${\cal R}_q\,$ assumes a finite matrix form
with entries expressed in terms of powers of $E, F, q^{\pm \frac{1}{2} H}\,,$
the latter generating a "double cover" ${\cal D}\,$ of $U_q\,$ (cf. the end of Section 2.2).
Evaluating, for example, all factors in (\ref{RUq}) in the fundamental
$2$-dimensional representation (\ref{Xf}), only the first two
terms of the sum survive, giving rise to the constant $4 \times 4\,$
matrix $R := {\cal R}_q^f = (R_{\a\b}^{\s\rho})$:
\be
R = q^{\frac{1}{2}} \left(\matrix{q^{-1}&0&0&0\cr 0&1&0&0\cr
0&-\l&1&0\cr 0&0&0&q^{-1}}\right)\ , \qquad{\rm or}\qquad
R_{\a\b}^{\s\rho} = q^{-\frac{1}{2}} \d_\a^\rho \d_\b^\s -
q^{\frac{1}{2}} \, {\cal E}^{\rho\s} {\cal E}_{\a\b}\ .
\lb{RRp}
\ee
The relations (\ref{aex2}) can be now reset as (homogeneous) $R$-matrix relations
\be
R({\hat p})\, a_1\, a_2 = a_2\, a_1 R \ ,\qquad{\rm i.e.}\qquad
R({\hat p})^{ij}_{\ell m} a^\ell_\a a^m_\b  = a^j_\rho a^i_\s R_{\a\b}^{\s\rho}
\lb{QMA2}
\ee
and a determinant condition,
\be
{\det}_q a := \frac{1}{[2]} \epsilon_{i\!j}\, a^i_\a a^j_\b\, {\cal E}^{\a\b} = [\hat p \, ]\ ,\qquad
(\epsilon_{i\!j} ) = \left(\matrix{0&-1\cr 1&0} \right)\ ;
\lb{detq2}
\ee
the {\em dynamical} $R$-matrix $R({\hat p})\,$ in (\ref{QMA2}) is given explicitly by
\be
\lb{RRp2}
\qquad R({\hat p})\, =\,
q^{\frac{1}{2}}
\left(\matrix{
q^{-1}&0&0&0\cr
0&{{[\hat p +1]}\over{[\hat p]}}&-\frac{q^{\hat p}}{[\hat p ]} &0\cr
0&\frac{q^{-\hat p}}{[\hat p]}&{{[\hat p -1]}\over{[\hat p ]} }& 0\cr
0&0&0&q^{-1}}\right) \ .
\ee
Both (\ref{QMA2}) and (\ref{detq2}) admit a straightforward generalization to arbitrary $n\,$ \cite{FHIOPT}.
It turns out that for $n=2\,$ the determinant condition alone implies (\ref{aex2}).

In the case of interest, when ${\cal A}_q\,$ appears as an ${\widehat{su}}(2)\,$ chiral zero modes' algebra,
$q\,$ is an even root of unity. If $h\,$ is the {\em height} of the ${\widehat{su}}(2)\,$ affine algebra
representation, equal to the sum of the (positive integer) level and the dual Coxeter number ($n\,,$
in the case of $su(n)$), then
\be
q^h = - 1\,,\qquad [2]= q+q^{-1} = 2 \cos \frac{\pi}{h}\,.
\lb{qh}
\ee
(To fit the expression (\ref{B1B2}) below for the braid group action on solutions of the
Knizhnik-Zamolodchikov equation, we will have to choose
$q=e^{-i\frac{\pi}{h}}\,$ as in \cite{FHIOPT, FHT6}. For our present purposes only the properties (\ref{qh}) will be needed.)
The QUEA $U_q\,$ admits for such $q\,$ a non-trivial ideal generated by $E^h\,,\, F^h\,$ and
$q^{2hH}-\id\,.$ Factoring $U_q\,$ by this ideal i.e., setting
\be
E^h = 0 = F^h\,,\qquad q^{hH} = q^{-hH} \ ,
\lb{Uqres}
\ee
we obtain the {\em restricted QUEA} $\bU\,$ (the "restricted quantum group" of \cite{FGST1}).
We shall see in Section 2.3 that the relations (\ref{Uqres}) are automatically satisfied in the
Fock space representation of ${\cal A}_q\,$ for $q\,$ satisfying (\ref{qh}).

As noted in the Introduction, the true counterpart of the extended chiral $\widehat{su}(2)_{h-2}\,$ WZNW model
is an (infinite dimensional) {\em extension} $\tU\,$ of $\bU\,$ which we proceed to define.

Introduce, following Lusztig (see \cite{L} and references therein), the "divided powers"
\be
E^{(n)} = \frac{1}{[n]!}\, E^n\,,\quad F^{(n)} = \frac{1}{[n]!}\, F^n\,,\quad [n] = \frac{q^n-q^{-n}}{q-q^{-1}}\,,\quad
[n]! = [n][n-1]!\,,\quad [0]! = 1
\lb{divpowEF}
\ee
satisfying
\ba
&&X^{(m)} X^{(n)} = \left[ {n+m}\atop{n}\right] X^{(m+n)}\,,\qquad X = E, F\ ,\nn\\
&&[ E^{(m)} , F^{(n)} ] = \sum_{s=1}^{min (m,n)} F^{(n-s)} \left[ {H+2s-m-n}\atop{s}\right] E^{(m-s)}\ ,
\lb{CRdp}
\ea
where the $q$-binomial coefficients\footnote{G. Lusztig
\cite{L} calls them "Gaussian binomial coefficients".} $\left[{a}\atop{b}\right]\,$
defined, for integer $a\,$ and non-negative integer $b\,,$ as
\be
\left[ {a\atop b}\right]:=
\prod_{t=1}^{b} \frac{q^{a+1-t} - q^{t-a-1}}{q^t-q^{-t}}\,, \qquad \left[ {a\atop 0}\right] := 1\ ,
\qquad\left[ {a\atop b}\right] \equiv \frac{[a]!}{[b]![a-b]!}\qquad {\rm for}\quad b\le a\ ,\quad
\lb{Gbinom}
\ee
are polynomials with integer coefficients of $q$ and $q^{-1}\,.$
The expressions (\ref{divpowEF}) are only defined for $m,n < h\,$ (as $[h]=0$); the relations (\ref{CRdp}),
however, make sense for all positive integers $m,n\,$ and can serve as an implicit definition of higher
divided powers. It is sufficient to add just $E^{(h)}\,$ and $F^{(h)}\,$ in order to generate
the extended $\tU\,$ algebra; their powers and products give rise to an infinite sequence of new elements --
in particular,
\be
\left( E^{(h)} \right)^n = \frac{[nh]!}{([h]!)^n}\, E^{(nh)} = \left( \prod_{\ell =1}^n
\left[ {\ell\, h}\atop{h} \right] \right)\, E^{(nh)} = (-1)^{\left(n\atop 2\right) h} n!\, E^{(nh)}\ .
\lb{Edpn}
\ee
To derive (\ref{Edpn}), we evaluate expressions of the type $\frac{[nh]}{[h]}\,$ as polynomials in $q^{\pm 1}\,$
and use the relation $[n h + m]=(-1)^n [m]\,$ to deduce
\be
\frac{[nh]}{[h]} = \sum_{\nu = 0}^{n-1} q^{(n-1-2\nu)h} = (-1)^{n-1}\, n\ ,\qquad\qquad
\left[ {n h}\atop{h} \right] = (-1)^{(n-1)\, h}\, n\ .
\lb{nhoverh}
\ee
The last result is a special case of the general formula \cite{L}
\be
\left[ {Mh+a\atop Nh+b}\right] = (-1)^{(M-1)Nh + aN - bM}\, \left[ {a\atop b}\right]\, \left( {M\atop N}\right)\,,
\lb{q-bin1}
\ee
valid for $q = e^{\pm\frac{i\pi}{h}}\,$ and $M\in {\Z}\,,\ N \in {\Z}_+\,,\ 0\le a,b\le h-1\,;$
here $\left( {M\atop N}\right) \in {\Z} \,$ is an {\em ordinary} binomial coefficient.

\medskip

\subsection{Monodromy and the quantum double}

The covariant chiral WZNW field $g(x) = \{ g^A_\a (x) \}\,$ can be written as a sum of tensor products
\be
g(x) = \sum_{i=1}^n u_i (x) \otimes a^i\,,\qquad u_i = \{\, u^A_i\,\}\,,\qquad a^i = \{\, a^i_\a\,\}\,,\qquad
A\,,\,\a = 1 , \dots , n\ .
\lb{g=uanew}
\ee
For $n=2\,,\  g(x)\,$ acts on a state space of the form
\be
{\cal H} = \oplus_{p=1}^\infty\, {\cal H}_p \otimes {\cal V}_p\qquad (\, {\dim} {\cal V}_p = p\, )\ .
\lb{HVp}
\ee
Here ${\cal H}_p\,$ is an (infinite dimensional) ${\widehat{su}}(2)_{h-2}\,$ current algebra module with a
$p$-fold degenerate ground state of (minimal) conformal energy $\Delta_p = \frac{p^2-1}{4h}\,,$ while ${\cal V}_p\,$ is
a ($p$-dimensional) $\bU\,$ module; $u_1\,$ and $a^1\,$ raise, $u_2\,$ and $a^2\,$ lower the weight $p\,$ by $1\,$ (in
particular, $u_2\, {\cal H}_1 = 0\,,\ a^2\, {\cal V}_1 = 0\,$). The field $g(x)\,$ and the {\em chiral vertex operator} $u(x)\,$
are multivalued functions of monodromy
\be
g(x+2\pi)  = g (x)\, M \,,\qquad u (x+2\pi) = M_p\, u (x)\ ,
\lb{gu-mon}
\ee
respectively, where $M_p\,$ is diagonal,
\be
(M_p)^i_j= q^{ 1 -\frac{1}{n}-2{\hat p}_i} \,\d^i_j \ , \quad q^{ 1 -\frac{1}{n}-2{\hat p}_i}
\, u_i = u_i\, q^{ \frac{1}{n}-1-2{\hat p}_i}\ ,\qquad  \prod_{i=1}^n q^{{\hat p}_i} = \id\,.
\lb{Mpn}
\ee
Thus the quantum matrix $a\,$ intertwines between the $U_q\,$ covariant monodromy $M\,$ and the diagonal one $M_p\,$
(\cite{HIOPT, FHIOPT}):
\be
a\, M = M_p\, a \ ,\qquad q^{1 -\frac{1}{n}-2{\hat p}_i }\, a^i= a^i\, q^{\frac{1}{n}- 1 -2{\hat p}_i }\ ;
\lb{Mpq}
\ee
in particular, for $n=2\,,$ we have
\be
M_p = q^{\frac{1}{2}}\, \left(\matrix{q^{-{\hat p}}&0\cr 0&q^{\hat p}\cr}\right)\ ,
\qquad p = p_{12}\equiv p_1-p_2 = 2p_1 \ .
\lb{Mp}
\ee
The braiding properties
\ba
&&g_1 (x)\, g_2 (y) = g_2 (y)\, g_1 (x)\, (R^-_{12}\, \theta (x-y) + R^+_{12}\, \theta (y-x) )\,,\qquad 0<|x-y|<2\pi\,,
\nn\\
&&R_{12}^- = R_{12}\,,\qquad R_{12}^+ = R_{21}^{-1}
\lb{braid-g}
\ea
require that the deformation parameter $q\,$ obeys (\ref{qh}) (see \cite{HST}).

The monodromy matrix $M\,$ admits a Gauss decomposition written in the form
\be
M = q^{\frac{1}{n}-n}\, M_+ M_-^{-1}\qquad ( \, \det M_+ = 1 = \det M_- \, )\,,
\lb{M}
\ee
where $M_+\,$ and $M_-\,$ are upper, resp. lower triangular matrices.
The Gauss components $M_\pm\,$ obey, as a consequence of (\ref{g=uanew}) and (\ref{braid-g}), the exchange relations
\be
(M_{\pm})_1\, a_2\, =\, a_2\, (R_{12}^{\pm})^{-1}\, (M_{\pm})_1\ ,
\lb{Mpmaq}
\ee
\be
R_{12} (M_\pm)_2 (M_\pm)_1 = (M_\pm)_1 (M_\pm)_2 R_{12}\ .
\lb{MpmM}
\ee
Eqs. (\ref{MpmM}) can be viewed (cf. \cite{FRT}) as the defining relations of a Hopf algebra $( U_q\bo_- \,, U_q\bo_+)\,$
compounded by two Borel algebras $U_q \bo_\pm\,$ corresponding to the entries of $M_\pm\,.$
Restricting attention to $n=2\,$ and setting
\be
\lb{qdouble}
M_- = \left(\matrix{ k_-& 0\cr \l\, k_-^{-1} E & k_-^{-1}\cr}\right)\,,\qquad
M_+  = \left(\matrix{ k_+^{-1}& -\l\, F\, k_+\cr 0 & k_+\cr}\right)\,,
\ee
we find, indeed, from (\ref{MpmM}) the characteristic relations for the pair of $q$-deformed Borel algebras,
\be
U_q\bo_-\, :\quad k_- E = q\, E \, k_- \,,\qquad
U_q\bo_+\,:\quad F k_+ = q\, k_+ F
\lb{B-ex}
\ee
and the mixed relations
\be
[ k_+ , k_- ] = 0\,,\quad k_+ E = q\, E\, k_+\,,\quad F\, k_- = q\, k_- F\,,\quad
[ E , F ] = \frac{k^{2}_- - k_+^{-2} }{\l}\ .
\lb{B-mix}
\ee
Applying the defining relations for the coproduct, the antipode and the counit
\be
\Delta (X^\a_{\b} ) = X^\a_{\s} \otimes X^\s_{\b} \,,\qquad S( X^\a_{\b} ) = (X^{-1})^\a_{\b}\,,\qquad
\e ( X^\a_{\b} ) = \d^\a_\b
\lb{coalg-m}
\ee
to $X=M_\pm\,,$ we find
\ba
&&\Delta (E) = E \otimes k_-^{2}+ \id\otimes E\,,\qquad\ \Delta (k_-) = k_-\otimes k_-\,,\nn\\
&&S(E) = - E\, k_-^{-2} \,,\quad\  S(k_-) = k_-^{-1}\,,\qquad
\varepsilon (E) = 0\,,\qquad\varepsilon (k_-) = 1\,,\lb{coalg-Chev-}\\
&&\Delta (F) = F \otimes \id + k_+^{-2}\otimes F \,,\qquad \Delta (k_+) = k_+ \otimes k_+\,,\nn\\
&&S(F) = - k_+^2 F \,, \qquad S(k_+) = k_+^{-1}\,,\qquad
\varepsilon (F) = 0\,,\qquad\varepsilon (k_+) = 1\,.\lb{coalg-Chev+}
\ea

We shall also consider the finite dimensional restricted algebra $( \bU \bo_-\,, \bU \bo_+ )\,$ by imposing the relations
\be
E^h = 0\,,\qquad k_-^{4h} = \id\,,\qquad F^h = 0\,,\qquad k_+^{4h} = \id\ .
\lb{bD}
\ee
We can view $( \bU \bo_-\,, \bU \bo_+ )\,$ as the (Drinfeld) {\em quantum double} regarding the
elements of $\bU \bo_-\,$ as linear functionals on $\bU \bo_+\,.$ (A similar interpretation of
the infinite dimensional algebra $( U_q\bo_- \,, U_q\bo_+ )\,$ would require topological considerations.)

The proof of the following propositions is analogous to those in \cite{FGST1}.

\medskip

\noindent
{\bf Proposition 2.1~} {\em Given the finite dimensional Hopf algebras $\bU \bo_\pm\,,$ there exists a unique
bilinear pairing $\left\langle Y  , X \right\rangle \ (\,\in {\C}\ $ for any $X\in \bU\bo_+\,,\, Y\in \bU\bo_-$)
such that, for $\Delta (X) = \sum_{(X)} X_1 \otimes X_2\,,$
\ba
&&\left\langle Y_1 Y_2 , X \right\rangle =
(Y_1 \otimes Y_2 )\, \Delta (X) \equiv
\sum_{(X)} \left\langle Y_1 , X_1 \right\rangle \left\langle Y_2 , X_2 \right\rangle\,,
\lb{kE1-0}\\
&&\left\langle \Delta (Y) , X_1\otimes X_2 \right\rangle \equiv \sum_{(Y)} \left\langle Y_1 , X_1 \right\rangle
\left\langle Y_2 , X_2 \right\rangle = \left\langle Y , X_2 X_1 \right\rangle \,,
\lb{kE2}\\
&&\left\langle \id , X \right\rangle = \e (X)\,,\qquad
\left\langle S(Y) , X \right\rangle = \left\langle Y , S^{-1}(X) \right\rangle\,,\qquad
\e (Y) = \left\langle Y , \id \right\rangle\ .\qquad
\lb{kE3}
\ea
It is given by
\be
\left\langle E^\mu k_-^m , f_{\nu n} \right\rangle = \d_{\mu\nu} \frac{[\mu ]!}{(-\l)^\mu}\,
q^{\frac{\mu(\mu-1)-mn}{2}}
\lb{kE}
\ee
where $\{ f_{\nu n} \}\,$ is a Poincar\'e-Birkhoff-Witt (PBW) basis in $\bU\bo_+\,:$
\be
f_{\nu n} := F^\nu k_+^n \ ,\qquad 0\le n\le 4h-1\,,\quad 0\le\nu\le h-1\,.
\lb{PBW-}
\ee
}

\noindent
{\bf Proposition 2.2~} {\em The mixed relations (\ref{B-mix}) are recovered provided the product $X Y\,$ is constrained by
\be
X\, Y (\bullet) = \sum_{(X)} Y(S^{-1} (X_3)\, {\bullet} \, X_1 )\, X_2
\lb{mult-pm}
\ee
for
\be
\Delta^{(2)} (X) =
(\id\otimes \Delta)\, \Delta (X) = ( \Delta\otimes\id)\, \Delta (X) =
\sum_{(X)} X_1\otimes X_2\otimes X_3\ .
\lb{D2}
\ee
In (\ref{mult-pm}) $Y(Z) \equiv \left\langle Y , Z \right\rangle\,$ and the dot $(\bullet )\,$
stands for the argument ( $Z\in \bU\bo_+\,$) of the functional.}

\medskip

The double cover ${\cal D}\,$ of the QUEA $U_q\,$ of Section 2.1 is obtained from $( U_q\bo_- \,, U_q\bo_+ )\,$
by imposing the relation
\be
k_- = k_+ =: k \qquad (\, q^H = k^2\,)\ ,
\lb{kpmk}
\ee
thus equating the diagonal entries of $M_+\,$ and $M_-^{-1}\,.$ Similarly, its {\em restriction} ${\overline{\cal D}}\,,$
the quotient of $( \bU \bo_-\,, \bU \bo_+ )\,$ for which both (\ref{bD}) and (\ref{kpmk}) hold so that, in particular,
\be
k^{4h} \equiv q^{2H} = \id\,,
\lb{rD}
\ee
is a double cover of $\bU\,.$

Eqs. (\ref{Mpmaq}) express the zero modes' covariance with respect to ${\cal D}\,$
(implying (\ref{Uqa21})).

\subsection{Fock space representation of ${\cal A}_q\,,\ \bU$ and $\tU$}

As already noted in Section 2.1, the algebra ${\cal A}_q\, $
is a $q$-deformation of Schwinger's oscillator algebra \cite{Schwinger}
giving rise to a model Fock space for the irreducible representations of $su(2)\,$ in which
$a^1_\a\,$ and $a^2_\a\,$ play the role of creation and annihilation operators, while the eigenvalues of ${\hat p}\,$
will be identified with the dimensions, $p=2I+1\,$ (for $I\,$ the "isospin"). We define the $U_q$-invariant vacuum state
$|1, 0{\cal i}\,$ by
\be
a^2_\a |1, 0{\cal i} = 0\,,\quad \a = 1,2\ ,\qquad X |1, 0{\cal i} = \e (X) |1, 0{\cal i}\qquad \forall X\in U_q\ .
\lb{vac}
\ee
The first relation in (\ref{vac}), together with (\ref{ap2}), (\ref{aex2}), requires
\be
(q^{\hat p}\, - q) |1,0{\cal i} = 0\ ,\qquad
a^2_\a a^1_\b |1, 0{\cal i}= {\cal E}_{\a\b} |1,0{\cal i}\ .
\lb{a2a1-0}
\ee
A basis $\{\, |p,m{\cal i}\,,\ p=1,2,\dots\,;\ 0\le m\le p-1\,\}$
in the Fock space ${\cal F}_q = {\cal A}_q\,|1,0{\cal i}\,$ is obtained by
acting on the vacuum by homogeneous polynomials (of degree $p-1\,$) of the creation operators $a^1_\a\,:$
\be
\lb{basis2}
|p,m{\cal i} := (a^1_1 )^m (a^1_2 )^{p-1-m} |1, 0{\cal i}\qquad\quad (\, (q^{\hat p} - q^p) |p,m{\cal i} = 0\, )\ .
\ee
The action of $a^i_\a\,$ on the basis vectors is given by
\ba
&&a^1_1 | p , m {\cal i} =  | p+1 , m+1 {\cal i} \,,\nn\\
&&a^1_2 | p , m {\cal i} =  q^m | p+1 , m {\cal i}\,,\nn\\
&&a^2_1 | p , m {\cal i} =  - q^{\frac{1}{2}} [p-m-1] | p-1 , m {\cal i}\,,\nn\\
&&a^2_2 | p , m {\cal i} =  q^{m-p+\frac{1}{2}} [m] | p-1 , m-1 {\cal i} \,.
\lb{apmn2}
\ea
Their $U_q\,$ properties follow from (\ref{basis2}), (\ref{divpowEF}), (\ref{Uqa21}) and (\ref{vac}):
\ba
&&q^H |p,m{\cal i} = q^{2m-p+1} |p,m{\cal i}\,,\nn\\
&&E^{(r)} |p,m{\cal i} = \left[{p-m-1}\atop{r}\right]\, |p,m+r{\cal i}\,,\nn\\
&&F^{(s)} |p,m{\cal i} = \left[{m}\atop{s}\right]\,\, |p,m-s{\cal i}\,.
\lb{Uqprop2}
\ea

We equip ${\cal F}_q\,$ with a symmetric bilinear form ${\cal h} ~~ | ~~{\cal i}\,$
introducing a "bra vacuum" ${\cal h}1, 0|\,$  dual to $| 1,0 {\cal i}\,$ such that
\be
{\cal h} 1, 0\, |\, a^1_\a = 0 = {\cal h} 1, 0 |\, (q^{\hat p} - q^p )\,,\qquad
{\cal h} 1, 0 | \, (X - \e (X)) = 0\quad \forall X\in U_q\,,\qquad
{\cal h} 1, 0 \, | \,1, 0 {\cal i} = 1\,,
\lb{bra-vac}
\ee
and a {\it transposition} (a linear antiinvolution) $A\ \rightarrow\ ^t\!\! A\,$ on ${\cal A}_q\,$
such that
\be
{\cal h} x |\, A\, y {\cal i}  \equiv
{\cal h} A\,y\,  | \, x\,{\cal i}= {\cal h}\, ^t\!\! A\, x\, |\, y\, {\cal i}\qquad
\forall\, x,y\in{\cal F}_q\,,\quad A\in {\cal A}_q\,.
\lb{transp}
\ee
The transposition defined on the ${\cal A}_q\,$ generators by
\be
\lb{transp2}
^t (q^{\hat p}) = q^{\hat p}\ ,\qquad\quad
^t (a^i_\a) = {\epsilon}_{i\! j}\, {\cal E}^{\a\b} a^j_\b\ ,
\ee
i.e. $^t (a^1_1) = q^{\frac{1}{2}} a^2_2\,,\
^t (a^1_2) = - q^{-\frac{1}{2}} a^2_1\,,$ and extended to products as an algebra antihomomophism,
$^t (A B) =\, ^t\! B \, ^t\! A\,,$ preserves the relations (\ref{ap2}), (\ref{aex2}). To verify this, as well as the
involutivity property $^t (^t\! A ) = A\,,$ one uses (\ref{eps1}), (\ref{eps2}) and the (undeformed) relation
${\epsilon}^{i s}{\epsilon}_{s j} = - \d^i_j\,.$
The relations (\ref{aex2}), (\ref{bra-vac}) and (\ref{transp2}) allow to compute the inner product of the basis vectors (\ref{basis2}):
\be
\lb{bilin2}
{\cal h} p, m| p', m'{\cal i} = \delta_{pp'} \delta_{mm'} q^{m(m+1-p)} [m]! [p-m-1]!\ .
\ee

Eq. (\ref{Mpq}) is easily verified to hold on ${\cal F}_q\,$ with $M_p\,$ given
by (\ref{Mp}) and $M\,$ obtained from (\ref{M}) (for $n=2\,$) and (\ref{qdouble}), (\ref{kpmk}), so that
\be
M = q^{-\frac{1}{2}}\, \left(\matrix{ \l^2\, F E + q^{-H-1}& - \l\, F q^{H-1}\cr - \l\,E & q^{H-1}\cr}\right)\ .
\lb{MpM}
\ee
The $U_q\,$ generators can be expressed, using (\ref{Mpq}) and (\ref{detq2}), in
terms of $a^j_\a\,,$ cf. \cite{FHIOPT}:
\ba
&&E = - q^{-\frac{1}{2}} a^1_1 a^2_1\,, \qquad
F q^{H-1} = q^{\frac{1}{2}}  a^1_2 a^2_2 = \, ^t\!E\,,\nn\\
&&q^H = q^{\frac{1}{2}} a^2_2 a^1_1 - q^{-\frac{1}{2}} a^1_1 a^2_2 =
q^{\frac{1}{2}} a^1_2 a^2_1 - q^{-\frac{1}{2}} a^2_1 a^1_2 =\, ^t(q^H )\, .
\lb{EFH}
\ea
Eq. (\ref{transp}) with $A = X\in U_q\,$ then follows, i.e. the bilinear form (\ref{bilin2}) is $U_q$-invariant.
The relations (\ref{EFH}) show that the action of the transposition
on the monodromy matrix is equivalent to the standard matrix transposition,
\be
^t (M^\a_\b) = M^\b_\a
\, .
\lb{transp-EFH}
\ee
\\

For generic $q\,,$ i.e. for $q\,$ not a root of unity, the $p$-dimensional space ${\cal V}_p\,$
spanned by $| p, m{\cal i}\,$ for $m= 0,\dots , p-1\,$ is an irreducible $U_q$ module and
\be
{\cal F}_q = \bigoplus_{p=1}^{\infty}\, {\cal V}_p
\lb{FV}
\ee
is a {\em model space} for $U_q\,$ in which every finite dimensional irreducible representation (IR)
appears with multiplicity one. (This result was established, more generally,
for the $U_q s\ell (n) \,$ Fock space in \cite{FHIOPT}.) The irreducible components of ${\cal F}_q\,$ are singled out by the
eigenvalues of the (rescaled) {\em Casimir operator}
\be
C = \l^2\, E F + q^{H-1} + q^{1-H} = \l^2\, F E + q^{H+1} + q^{-H-1}\ .
\lb{C}
\ee
On the Fock space we have
\be
(C - q^{\hat p}-q^{-{\hat p}} )\, {\cal F}_q = 0\ .
\lb{CF}
\ee

For $q\,$ satisfying (\ref{qh}) and $p > h\,,\ {\cal V}_p\,$ carries an indecomposable representation of $\bU\,$ --
it admits $\bU$-invariant subspaces with no invariant complements.

In order to describe its structure we first
observe that $\bU\,$ has exactly $2h\,$ IRs $\, V_p^\pm\,,\ 1\le p\le h\,$ defined as eigensubspaces of the
operator $q^{\hat p}\,:$
\be
\lb{2h-IR}
(q^{\hat p} - \epsilon\, q^p)\, V_p^\epsilon = 0\qquad(\,{\rm dim}\, V_p^\epsilon = p\, )\,, \qquad 1\le p\le h\ ;
\ee
we shall refer to the sign $\epsilon\,$ as to the {\em parity} of the IR $V_p^\epsilon\,.$
The weight basis $| p, m {\cal i}\,,\ \ 0\le m\le p-1\,$ (\ref{Uqprop2}) of $V_p^+\,$
can be extended to $V_p^-\,$ so that
\be
q^H | p, m {\cal i}^\epsilon = \epsilon \,  q^{2m-p+1} | p, m {\cal i}^\epsilon \,,\qquad
E\, | p, p-1 {\cal i}^\epsilon  = 0 = F\, | p, 0 {\cal i}^\epsilon  \ .
\lb{Uqres-EFK}
\ee
Noting the relations
\be
( EF - [N_+ ] [N_- +1] ) \, {\cal F}_q = 0 = ( FE - [N_+ +1] [N_-] )\, {\cal F}_q\,,
\qquad N_\pm := \frac{1}{2}\, ({\hat p} -1 \pm H)\
\lb{Npm}
\ee
along with (\ref{Uqres-EFK}), we find
\be
(EF - \epsilon \,[m][p-m] )\, | p, m {\cal i}^\epsilon  = 0 =(FE - \epsilon \,[m+1][p-m-1] )\, | p, m {\cal i}^\epsilon \,.
\lb{EFpm}
\ee

\medskip

\begin{figure}[htb]
\centering
\includegraphics*[bb=10 190 580 620,width=\textwidth]{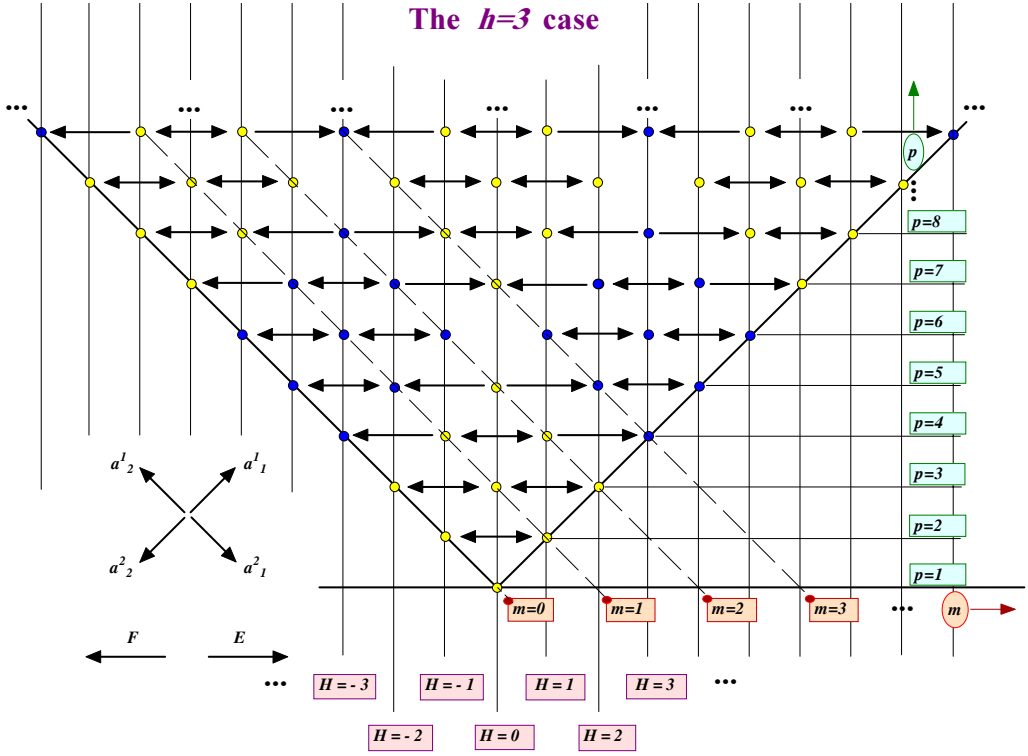}
\caption{\scriptsize{The $\bU$ representation on the Fock space ${\cal F}_q$ for $q= e^{\pm i\frac{\pi}{3}}\,.$
Vectors belonging to (sub)modules or subfactors of type $V_p^+\,$ (for some $p\,$)
are denoted by
${\circ}\,,$ and those belonging to $V_p^-\,$ -- by
${\bullet}\,.$
}}
\end{figure}

\medskip

The {\em negative parity} representations $V_p^-\,,\ 1\le p\le h\,$
first appear as subrepresentations of the Fock space modules
${\cal V}_{h+p}\,$ that admit two invariant submodules isomorphic to
them (spanned by $\{ | h + p , m {\cal i} \}\,$
and $\{ | h + p , h+ m {\cal i} \}\,$ for $m=0,\dots , p - 1\,$);
they both obey (\ref{Uqres-EFK}), (\ref{EFpm}), albeit $E\,$ and $F\,$ act
differently, their actions being related by an equivalence transformation:
\ba
&&E | h + p , m {\cal i} = - [p-m-1]\, | h + p , m+1 {\cal i}\,,\nn\\
&&F | h + p , m {\cal i} = [m]\, | h + p , m-1 {\cal i}\,,
\lb{Vp1}\\ \nn\\
&&E | h + p , h+ m {\cal i} \} = [p-m-1]\, | h + p , h+ m+1 {\cal i} \}\,,\nn\\
&&F | h + p , h+ m {\cal i} \} = - [m]\, | h + p , h+ m-1 {\cal i} \}\,,
\qquad 0\le m\le p-1
\lb{Vp2}
\ea
(we identify $| p , m {\cal i}^-\,$ with
either $| h + p , h+ m {\cal i}\,$ or $(-1)^m | h + p , m {\cal i}\,$).
For $p = h\,$ these two subrepresentations exhaust the content of
${\cal V}_{h+p}\,:\ \ {\cal V}_{2h} = V^-_h \oplus V_h^-\,.$
For $1\le p \le h-1\,$ the quotient of ${\cal V}_{h+p}\,$ by the direct sum of
invariant subspaces is isomorphic to $V_{h-p}^+\,.$ Thus, the subquotient
structure of ${\cal V}_{h+p}\,$ is described by the short exact sequence
\be
0\ \rightarrow\ V^-_p \oplus V^-_p\ \rightarrow\ {\cal V}_{h+p}\
\rightarrow V^+_{h-p}\ \rightarrow\ 0\ .
\lb{shexseq}
\ee
More generally (cf. Figure 1), the structure of ${\cal V}_{Nh+p}\,$
as a $\bU\,$ module can be described by the short exact sequence
\ba
&&0\ \ \rightarrow\ \underbrace{V^{\epsilon (N)}_p \oplus V^{\epsilon (N)}_p\dots \oplus V^{\epsilon (N)}_p}
\ \ \rightarrow\ \ {\cal V}_{Nh+p}\ \
\rightarrow\ \  \underbrace{V^{-\epsilon (N)}_{h-p}\oplus \dots \oplus V^{-\epsilon (N)}_{h-p}}\ \rightarrow\ 0\nn\\
&& \hspace{20mm} \#\, (N+1)
\hspace{53mm}\#\, N
\lb{shexseqN}
\ea
(we have $N+1\,$ submodules $V^{\epsilon (N)}_p\,$ and a quotient module which is a direct sum of $N\,$ copies of
$V^{-\epsilon (N)}_{h-p}\,$), where $\epsilon (N)\,$ coincides with the parity of $N\,$ and $V^\pm_0\,$
consist of the $0\,$ vector:
\be
\epsilon (N)\, =\, (-1)^N\,,\qquad  V^\pm_0 = \{ 0 \} \ .
\lb{epN}
\ee

For $N\ge 2\,,\ 1\le p\le h-1\,,$ the indecomposable $\bU\,$ modules ${\cal V}_{Nh-p}\,$ are equivalent to the
"$M$-modules" ${\cal M}^{\epsilon (N)}(N)\,$ introduced in Section 1.6 of \cite{FGST2} (see also Section 3.2
of \cite{S2})\footnote{The authors thank A.M. Semikhatov for pointing out this relation to them.}; note that
the "parity" of the corresponding $M$-module is that of $N\,$.

The representations of the extended QUEA $\tU\,$ in ${\cal F}_q\,$ are easily
described on the basis of the above analysis.

\medskip

\noindent
{\bf Proposition 2.3~} {\em

\noindent
(a) The irreducible $\bU\,$ modules $V^+_p\ (1\le p\le h )$ extend to
$\tU$ modules, with $E^{(h)}\,$ and $F^{(h)}\,$ acting trivially.

\noindent
(b) The fully reducible $\bU$ modules ${\cal V}_{Nh}\,$ extend to irreducible
$\tU\,$ modules.

\noindent
(c) The structure of the extended $\tU\,$ modules ${\cal V}_{Nh+p}\,$
for $1\le p\le h-1\,,\ N=1,2,\dots\,$ is again given by the short exact sequence
(\ref{shexseqN}) but with the direct sums viewed
as irreducible representations of $\tU\,:$
\be
0\ \ \rightarrow\ V^{\epsilon (N)}_{N+1 ,\, p} \ \ \rightarrow\ \ {\cal V}_{Nh+p}\ \
\rightarrow\ \  V^{-\epsilon (N)}_{N ,\, p}\ \ \rightarrow\ \ 0
\lb{shex-eqN}
\ee
where
\be
V^{\epsilon (N)}_{N+1 ,\, p} := \oplus^{N+1} V_p^{\epsilon (N)}=\oplus_{n=0}^N\
Span \, \{\, |Nh+p , nh+m {\cal i}\,\}_{m=0}^{p-1}
\lb{VNp}
\ee
and $V^{-\epsilon (N)}_{N,\, p} = \oplus^N V^{-\epsilon (N)}_{h-p}\,$ are both
irreducible with respect to $\tU\,.$
}
\medskip

\noindent
{\bf Proof~} Using (\ref{Uqprop2}) and the relation $\left[{n}\atop h\right]=0\,$
for $n<h\,,$ we find
\be
E^{(h)} | p , m {\cal i} \, =\, 0\, =\, F^{(h)} | p , m {\cal i}\qquad{\rm for}
\quad p\le h\ ,
\lb{EhFhzero}
\ee
proving (a). On the other hand, $E^{(h)}\,$ and $F^{(h)}\,,$ shifting the label $m\,$ by $\pm h\,$
combine, for $N\ge 1\,,$ otherwise disconnected (equivalent) irreducible $\bU\,$ submodules of subquotients into a
single irreducible representation of $\tU\,:$  the relation
\ba
&&E^{(h)} | Nh+p , nh+m{\cal i} = \left[{(N-n)h+p-m-1}\atop{h} \right]\,
| Nh+p , (n+1)h+m{\cal i} =\nn\\
&&= (-1)^{(N-n-1)h+p-m-1} \, (N-n)\, | Nh+p , (n+1)h+m{\cal i}
\lb{Eh1}
\ea
where $n=0,1,\dots ,N-1\,,\ 1\le p \le h\,,\ 0\le m\le p-1\,,$ and a similar relation involving $F^{(h)}\,,$
imply (b), for $p=h\,,$ and the first part of (c), for $p<h\,.$
(The $q$-binomial coefficient in the second Eq. (\ref{Eh1}) is a special case of the
general formula (\ref{q-bin1}).)
The second part of (c) involving $V^{-\epsilon (N)}_{N, p}\,$ is obtained using
\be
E^{(h)} | Nh+p , nh+p+m{\cal i} = \left[{(N-n)h-m-1}\atop{h} \right]\,
| Nh+p , (n+1)h+p+m{\cal i}
\lb{Eh2}
\ee
for $N\ge 1\,,\ n=0,1,\dots ,N-2\,,\ 1\le p \le h-1\,,\ 0\le m\le h-p-1\,$
(and a similar relation for $F^{(h)}$). \eod

\medskip

A (partial) information about an indecomposable representation is its content
in terms of irreducible modules, independently of whether they appear as its submodules or subquotients.
It is captured by the concept of the Grothendieck ring.
We write $R = R_1 + R_2\,$ if one of the representations in the right hand
side is a subrepresentation of $R\,$ while the other is the corresponding quotient representation,
and complete the structure to that of an abelian group by introducing formal differences
(so that e.g. $R_1 = R - R_2\,$) and zero element, given by the vector $\{ 0 \}\,.$
To define the GR multiplication, we start with the tensor product of
irreducible representations defined by means of the coproduct,
\be
(R_1 \otimes R_2 ) (x) = \sum_{(x)} R_1(x_1) \otimes R_2(x_2)\,,\qquad
x\in\bU\,,\quad \Delta (x) = \sum_{(x)} x_1 \otimes x_2
\lb{tens-ring}
\ee
and further, represent each of the (in general, indecomposable) summands in the expansion
by the GR sum of its irreducible submodules and subquotients (thus "forgetting" its indecomposable structure).

In the case of the restricted QUEA $\bU\,$ the GR is
the commutative ring ${\mathfrak S}_{2h}\,$ generated by the $2h\,$
irreducible representations $V_p^{\pm}\,,\ 1\le p\le h\,,$ while the GR for $\tU\,$ in ${\cal F}_q\,$
is generated by the irreducible representations $V^+_p\,,\ V^{\epsilon (N)}_{N+1 ,\, p}\,$
and $V^{-\epsilon (N)}_{N ,\, p}\,$ for $1\le p\le h\,,\ N=1,2,\dots .$
The GR content of ${\cal V}_{Nh+p}\,$ for $1\le p\le h\,,\quad N\in {\Z}_+\,$
which replaces the precise indecomposable structure given in Eqs. (\ref{shexseqN}) and (\ref{shex-eqN}) is
\be
{\rm GR}\, ( \bU )\,:\quad{\cal V}_{Nh+p} = (N+1)\, V^{\epsilon(N)}_p + N\, V^{-\epsilon(N)}_{h-p}
\lb{GRpb}
\ee
and
\be
{\rm GR}\, ( \tU )\,:\quad{\cal V}_{Nh+p} =  V^{\epsilon(N)}_{N+1 ,\, p} +  V^{-\epsilon(N)}_{N ,\, h-p}\quad ,
\lb{GRpt}
\ee
respectively. Note that any ${\cal V}_{Nh+p}\,$ contains an odd number
of irreducible $\bU\,$ modules of type $V^+\,$ and an even number of modules of type $V^-\,.$
The same "parity rule" is respected by the decomposition of the
$\tU\,$ IRs, described above, in terms of $V^\pm\,.$

Although all IRs of $\bU\,$ are contained in ${\cal F}_q\,,$ the restricted QUEA $\bU\,$
is not represented faithfully in our Fock space. As we shall see in Section 3 below,
the expression (\ref{CF}) for the Casimir operator $C\,$ in terms of
$q^{\hat p}+q^{-{\hat p}}\,$ on ${\cal F}_q\,$ implies that the radical of the centre
${\cal Z}_q\,$ of $\bU\,$ is represented trivially on the Fock space. Eq.
(\ref{CF}) together with the first equation (\ref{Uqprop2}) allows, on the other hand, to
express the central element $q^{hH}\,$ of $\bU\,$ as a polynomial of degree $h\,$ in $C\,.$
Indeed, the easily verifiable relations
\be
(q^{hH}+q^{h{\hat p}})\, {\cal F}_q = 0 = (q^{h{\hat p}}-q^{-h{\hat p}})\, {\cal F}_q
\lb{hHp}
\ee
imply
\be
\left( q^{hH} + \frac{1}{2}\, (q^{h{\hat p}}+q^{-h{\hat p}})\right) {\cal F}_q =
\left( q^{hH} + T_h (\frac{C}{2} ) \right) {\cal F}_q = 0\,,
\lb{hHCheby}
\ee
where $T_m(x)\,$ is the Chebyshev polynomial of the first kind (${\rm deg}\, T_m = m$) defined by
\be
T_m\, (\cos t)  =  \cos m\, t\ .
\lb{Cheby1}
\ee
It is all the more remarkable that the equation $q^{hH} + T_h (\frac{C}{2} ) = 0\,$
is valid algebraically i.e., not just when applied to ${\cal F}_q\,$ -- see Eq. (\ref{qhHTh1}) in Section 3.2 below.

We end up this section by describing the structure of ideals (and quotients) of
the restricted quantum matrix algebra (\ref{extA}) and its extension
${\tilde{\cal A}}_q\supset\tU\,.$ They both admit a sequence of nested ideals
\be
{\cal I}_h\, \supset\, {\cal I}_{2h} \,\supset\dots
\lb{nest-id}
\ee
where ${\cal I}_{Nh}\,$ is generated by all products of the form $(a_\b^i )^{h\nu}(a_\gamma^j )^{h(N-\nu )}\,,\ \nu=0,1,\dots ,N\,.$
The factor algebras ${\cal A}_{Nh} := {\cal A}_q / {\cal I}_{Nh}\,,$
\be
{\cal A}_h\,\subset \,{\cal A}_{2h} \,\subset\dots
\lb{nest-fac}
\ee
(the inclusions in (\ref{nest-fac}) are opposite to those of (\ref{nest-id})) are all finite dimensional.
We have considered in our earlier work (see, e.g. \cite{Goslar}) the corresponding
$h^2$-dimensional Fock space ${\cal F}_h = {\cal A}_h | 0 {\cal i}\,$ which
only involves the irreducibles representations $V^+_p\,$ of $\bU\,$
and does not admit a non-trivial extension to $\tU\,.$

\medskip

\section{The centre and the fusion ring of $\bU\,$ and of its Lusztig extension $\tU\,$}

\setcounter{equation}{0}
\renewcommand\theequation{\thesection.\arabic{equation}}

In this section we describe:
\newline
(1) the pair $( \bU\,, \bD )\,$ as finite dimensional
(factorizable and quasitriangular, respectively) Hopf algebras;
\newline
(2) the centre ${\cal Z}_q\,$ of $\bU\,$
and its relation to the GR of $\bU\,$ and $\tU\,.$
\newline
To make the exposition self-contained, we have put
together some basic facts and results of \cite{Dr, RS, Ker, Sch, FGST1} (using our conventions),
completing occasionally the arguments. Theorem 3.1 and the proofs of Propositions 3.1 and 3.5 are new.

\subsection{$\bU\,$ as a factorizable Hopf algebra. The Drinfeld map}

We begin by recalling the construction \cite{FGST1} of $\bU\,$ as a factorizable Hopf algebra.
To begin with, the finite dimensional quantum double $( \bU \bo_-\,, \bU \bo_+ )\,$
possesses an universal $R$-matrix given by the standard formula
\be
{\cal R}^{double}=
\sum_{\nu=0}^{h-1} \sum_{n = 0}^{4h-1} f_{\nu n} \otimes e_{\nu n}\,
\lb{Rdouble}
\ee
where $f_{\nu n}\,$ is defined by (\ref{PBW-}) and
\be
e_{\mu m} = \frac{(-\l)^\mu q^{-\frac{\mu (\mu -1)}{2}}}{4h\,[\mu ]!}
\sum_{r=0}^{4h-1} q^{\frac{mr}{2}}\, E^\mu k_-^r \,,\qquad\quad
\left\langle e_{\mu m} , f_{\nu n}   \right\rangle = \d_{\mu\nu} \d_{mn} \,,
\lb{dual+}
\ee
form dual PBW bases of  $\bU \bo_-\,$ and $\bU \bo_+\,,$ respectively (the prefactor in $e_{\mu m}\,$
being fixed by (\ref{kE})).

Let ${\mathfrak A}\,$ be an almost cocommutative Hopf algebra; given
the universal $R$-matrix, we can always construct the (universal)
{\em $M$-matrix} $\,{\cal M}\,$ that commutes with the coproduct,
\be
{\cal M} := {\cal R}_{21} {\cal R}
= \sum_{i} m_i\otimes m^i\, \in {\mathfrak A}\otimes {\mathfrak A}\qquad\Rightarrow\qquad
{\cal M}\, \Delta (x) = \Delta (x) {\cal M}\ .
\lb{M-matr}
\ee
A Hopf algebra is called {\em factorizable}, if both $\{ m_i \}\,$ and $\{ m^i \}\,$ form bases of it;
a finite dimensional quantum double is always factorizable \cite{RS}. (The opposite extreme is the case
of {\em triangular} Hopf algebra for which ${\cal R}_{21} = {\cal R}^{-1}
\,$ and hence, ${\cal M} = \id\otimes\id\,.$)

From (\ref{Rdouble}) and (\ref{dual+}) one readily obtains the $R$-matrix for the quotient $\bD\,$
obtained by the identification $k_\pm \equiv k\,$ (\ref{kpmk})
(with $k^{4h} = \id\,$):
\be
{\cal R} = \frac{1}{4h}\, \sum_{\nu = 0}^{h-1}
\frac{q^{-\frac{\nu (\nu -1)}{2}} (-\l)^\nu}{[\nu ]!}\,
F^\nu\otimes E^\nu \sum_{m ,\,n=0}^{4h-1} q^{\frac{mn}{2}} k^m \otimes k^n  \ \in\, \bD\otimes\bD\,.
\lb{RbD}
\ee
It is easy to see that, evaluating the universal $R$-matrix (\ref{RbD})
in the tensor square of the two-dimensional
representation (\ref{Xf}) (for $k=q^{\frac{H}{2}}\,$), one obtains
\ba
&&{\cal R}^f \equiv (\pi_f\otimes \pi_f )\, {\cal R} =  \frac{1}{4h}\,
\left( \id_2 \otimes \id_2 - \l\, F^f \otimes E^f \right)\,
\sum_{m ,\, n=0}^{4h-1} q^{\frac{mn}{2}} q^{m\frac{H^f}{2}}\otimes q^{n\frac{H^f}{2}} =\nn\\
&&= \left(\matrix{1&0&0&0\cr
0&1&0&0\cr
0&-\l&1&0\cr
0&0&0&1} \right) \left(\matrix{
q^{-\frac{1}{2}}&0&0&0\cr
0&q^{\frac{1}{2}}&0&0\cr
0&0&q^{\frac{1}{2}}&0\cr
0&0&0&q^{-\frac{1}{2}}} \right)
= q^{\frac{1}{2}}\,\left(\matrix{q^{-1}&0&0&0\cr
0&1&0&0\cr
0&-\l&1&0\cr
0&0&0&q^{-1}} \right) \ ,
\lb{Rf}
\ea
which coincides with $R\,$ of (\ref{RRp}); one uses the summation formula
\be
\sum_{m=0}^{4h-1} q^{\frac{mj}{2}} =
\left\{
\begin{array}{ll}
4h \ &{\rm for}\ j\equiv 0\ mod\ 4h\\
0 \ & {\rm otherwise}
\end{array}
\right.\ .
\lb{sum-m}
\ee

We shall also give, for completeness, the formula for the finite dimensional counterpart of (\ref{RUq21-1}):
\be
{\cal R}_{21}^{-1} = \frac{1}{4h}\, \sum_{m ,\,n=0}^{4h-1} q^{-\frac{mn}{2}} k^m \otimes k^n \sum_{\nu = 0}^{h-1}
 \frac{q^{\frac{\nu (\nu -1)}{2}} \l^\nu}{[\nu ]!}\,
E^\nu\otimes F^\nu  \ .
\lb{RbD21-1}
\ee
Note that ${\cal R}^f\,$ and $({\cal R}_{21}^{-1})^f\,$ are of {\em opposite}
triangularity.

The restricted QUEA $\bU\,$ is the Hopf subalgebra of $\bD\,$ generated by $E\,,\, F\,$ and $q^H = k^2\,.$
Its dimension is $2h^3\,,$ a PBW basis being provided e.g. by the elements
\be
\{ E^\mu F^\nu q^{n H}\,,\ 0\le \mu ,\nu \le h-1\,,\ 0\le n\le 2h-1\}\,.
\lb{PBW-Uqres}
\ee
Clearly, $\bU\,$ is not even almost cocommutative, since ${\cal R}\,$ (\ref{RbD})
does not belong to its tensor square.
Remarkably however, the expression for the corresponding $M$-matrix obtained from (\ref{RbD})
\be
{\cal M} =\frac{1}{2h}\,\sum_{\mu ,\nu = 0}^{h-1}
\frac{(-\l)^{\mu+\nu}q^{\frac{\nu(\nu+1)-\mu(\mu-1)}{2}}}{[\mu ]![\nu ]!}\,
\sum_{m,\,n = 0}^{2h-1} q^{mn +\nu(n-m) }
E^\mu F^\nu q^{mH} \otimes F^\mu E^\nu q^{nH}\ .
\lb{Mmatr}
\ee
only contains even powers of $k\,$ and hence, belongs to
$\bU \otimes \bU\,$ \cite{FGST1}\footnote{Note that, to comply with
our previous conventions for the zero modes, we have chosen here the
"dual" Drinfeld double with respect to the one in \cite{FGST1}, keeping the same Hopf structure for $\bU\,.$
In effect, our universal $R$-matrix (\ref{RbD}) coincides with ${\cal R}_{21}^{-1}\, $ of \cite{FGST1} and hence,
the $M$-matrix (\ref{Mmatr}) is the {\em inverse} of the one given by Eq. (4.4) of \cite{FGST1}
(in which there is a wrong $q\,$ factor that, happily, does not affect the computation of the Drinfeld images (4.6)).}.
Moreover, it is of the form
${\cal M} = \sum_{i=1}^{2h^3} m_i\otimes m^i\,$ where
$\{ m_i\}\,$ and $\{ m^i\}\,$ are two {\em bases} of $\bU\,,$ and
the latter fact implies that $\bU\,$ is factorizable, while its
quasitriangular "double cover" $\bD\,$ is not.

The relation between the "universal" $M$-matrix (\ref{Mmatr}) and the $2\times 2$
monodromy matrix with operator entries $M\,$ (\ref{MpM}) is simple and quite natural.
Computing $(\pi_f\otimes id)\,{\cal M}\,,$ we get (by taking first the sums in $m\,$)
\be(\pi_f\otimes id)\,{\cal M}
= \frac{1}{2h}\,\sum_{m,\,n=0}^{2h-1}\,
\left(\matrix{{(q^{m(n+1)} + \l^2 q^{mn+n+1} FE)\, q^{nH}}& {-\l\, q^{m(n-1)} F q^{nH}}\cr
{-\l\, q^{mn+n+1} E q^{nH}} & {q^{m(n-1)} q^{nH}} }\right)\nn\\
= q^{\frac{3}{2}} M\ .
\lb{calcM}
\ee

The inverse $M$-matrix ${\cal M}^{-1} = {\cal R}^{-1} {\cal R}_{21}^{-1}\,$ can be considered as
the monodromy associated to the alternative choice (${\cal R}_{21}^{-1}\,$) for the $R$-matrix.

Suppose that ${\mathfrak A}\,$ is a finite dimensional Hopf algebra
(such that an $M$-matrix ${\cal M}\in{\mathfrak A}\otimes{\mathfrak A}\,$ exists),
and let ${\mathfrak A}^*\,$ be its linear dual. The importance of the map
\be
\hat D:\ {\mathfrak A}^*\ \rightarrow\ {\mathfrak A}\,,\qquad \phi\ \mapsto\ (\phi\otimes id ) ({\cal M}) \equiv
\sum_{(m)} \phi(m_1)\, m_2\qquad \forall\,\phi\in {\mathfrak A}^*
\lb{Dr-map}
\ee
(called the {\em Drinfeld map} in \cite{FGST1}) has been clarified in \cite{Dr}.
Factorizable Hopf algebras are those for which $\hat D\,$ is a linear isomorphism, so that
$\hat D({\mathfrak A}^*) = {\mathfrak A}\,$ and $\hat D\,$ is invertible (the equivalence with the previous
definition is a simple exercise of linear algebra).

The space of {\em ${\mathfrak A}$-characters}
\be
{\mathfrak C}{\mathfrak h}  := \{\, \phi \in {\mathfrak A}^*\, |\ \phi (x y) = \phi (S^2(y) x)\ \ \forall\, x,y\in {\mathfrak A} \}
\lb{Ch-Ad*inv}
\ee
is an algebra under the multiplication defined by $(\phi_1 .\, \phi_2) (x) = (\phi_1\otimes\phi_2)\, \Delta (x)\,$
(for ${\mathfrak A}\,$ quasitriangular, this algebra is commutative \cite{Dr}).
Denote by ${\cal Z}\,$ the centre of ${\mathfrak A}\,,$ and by ${\mathfrak A}^\Delta\,$ the subalgebra of
${\mathfrak A}\otimes {\mathfrak A}\,$ consisting of
elements $d\,$ such that $[ d\, ,\, \Delta (x) ] = 0\quad \forall x\in {\mathfrak A}\,.$
It has been proven by Drinfeld (Proposition 1.2 of \cite{Dr}) that
\be
\phi \in {\mathfrak C}{\mathfrak h}\,,\quad d\in {\mathfrak A}^\Delta\,\qquad \Rightarrow\qquad
(\phi\otimes id ) (d) \in {\cal Z}\ .
\lb{Ch-AD-Z}
\ee
Since ${\cal M}\in {\mathfrak A}^\Delta\,,$ cf. (\ref{M-matr}), $\hat D\,$ also sends ${\mathfrak A}$-characters to central elements;
more than that, the restriction of the Drinfeld map on the ${\mathfrak A}$-characters
has the special property to provide a {\em (commutative)  algebra homomorphism} ${\mathfrak C}{\mathfrak h}\
\rightarrow\ {\cal Z}\,$ (Proposition 3.3 of \cite{Dr}),
\be
\hat D (\phi_1 . \, \phi_2 ) =  \hat D (\phi_1 ) \, \hat D ( \phi_2 )\qquad \forall\, \phi_1\,,\, \phi_2 \in {\mathfrak C}{\mathfrak h}
\lb{D-homom}
\ee
which, for ${\mathfrak A}\,$ factorizable, is in fact an isomorphism (Theorem 2.3 of \cite{Sch}).

In this case we have an alternative description of the space
of characters in terms of more tractable objects -- the elements of the centre ${\cal Z}\,.$

\medskip

\subsection{ The centre ${\cal Z}_q\,$ of $\bU\,$ and its semisimple part}

The restricted QUEA $\bU\,$ has a $(3h-1)$-dimensional centre ${\cal Z}_q\,,$ cf. \cite{FGST1},
which we proceed to describe, starting with the algebra of the rescaled Casimir operator (\ref{C}).
The following Proposition provides a compact expression for the central element $q^{hH}\,,$
see (\ref{qhHTh1}) (equivalent to (3.6) of \cite{FGST1} given there without derivation),
as well as a proof of (\ref{P2h=0}) which only uses the defining relations of $\bU\,.$

\medskip

\noindent
{\bf Proposition 3.1~} {\em

\noindent
(a) The central element $q^{hH}\,$ is related to $C\,$ by
\be
q^{hH} = - T_h (\frac{C}{2})\ ,
\lb{qhHTh1}
\ee
where $T_h\,$ is the $h$-th Chebyshev polynomial of the first kind (\ref{Cheby1}).

\noindent
(b)
The commutative subalgebra of $\,\bU\,$ generated by $C\,$ is $2h$-dimensional,
the charac-teristic equation of $C\,$ being
\be
P_{2h}(C) := \prod_{s=0}^{2h-1} (C - \b_s )= 0\ , \qquad \b_s = q^s + q^{-s} = 2\,\cos \frac{s\pi}{h}\ .
\lb{P2h=0}
\ee}
\medskip

\noindent
{\bf Proof~}
We shall start by writing the formula (see, e.g., $1.395$ in \cite{GR})
\be
\cos N t - \cos N y = 2^{N-1} \prod_{s=0}^{N-1} ( \cos t - \cos (y+\frac{2\pi s}{N} ) )
\lb{flaRG1}
\ee
for $2 \cos t = C\,$ (\ref{C}) and $e^{iy} =: Z\ $ (such that $Z^{2N} = 1)\,$ and applying it to the case
when $C\,$ and $Z\,$ are commuting operators in a finite dimensional space. We find
\be
2\, (T_N(\frac{C}{2}) - Z^N ) = \prod_{s=0}^{N-1} ( C - Z e^{\frac{2\pi i s}{N}} - Z^{-1} e^{-\frac{2\pi i s}{N}} )
\qquad{\rm for}\qquad Z^{2N} = \id\,.
\lb{TNZ}
\ee
Two special cases of (\ref{TNZ}): i)\ $N=2h\,,\ Z= \id\,$ and ii)\ $N=h\,,\ Z = q^{H-1}\,$
(for $q\,$ obeying (\ref{qh}) and $q^{2hH} = \id\,$) give
\be
2\, (\, T_{2h} (\frac{C}{2}) - \id ) = P_{2h} (C)
\lb{P2hT2h}
\ee
and
\be
2\, (\, T_h (\frac{C}{2}) + q^{hH} ) = \prod_{s=0}^{h-1} (C-q^{H-2s-1}-q^{-H+2s+1})\,,
\lb{qhHTh}
\ee
respectively. The following relations can be easily proved by induction in $r\,:$
\be
\l^{2r} E^r F^r = \prod_{s=0}^{r-1} (C - q^{H-2s-1} - q^{-H+2s+1} )\,,\qquad
\l^{2r} F^r E^r = \prod_{s=0}^{r-1} (C - q^{H+2s+1} - q^{-H-2s-1} )\, .
\lb{ErFr}
\ee
Setting $r=h\,$ and using (\ref{Uqres}), we deduce that the product in (\ref{qhHTh}) vanishes, proving (a).
Further, since $T_{2m} (\cos t ) = \cos 2 m t = 2\, (T_m (\cos t))^2 -1\,,$
(b) follows from (\ref{P2hT2h}) and (\ref{qhHTh1}):
\be
P_{2h}(C)\,= 4\, (\, q^{2hH} - \id ) = 0\ .
\lb{directP2h}
\ee
Hence, $P_{2h}\,$ is indeed the characteristic polynomial of $C \in {\cal Z}_q\,.$
\eod

\medskip

Since $\b_{2h-p} = \b_p\,$ and $\b_p \ne \b_r\,,\ 0\le p\ne r \le h\,,$ there are only $h+1\,$
different characteristic numbers $\b_s\,$ in (\ref{P2hT2h}); noting
that $\b_0 = 2 = -\b_h\,,$ one can write
\be
P_{2h} (x) = (x^2-4) \prod_{p=1}^{h-1} (x - \b_p )^2 \equiv (x^2-4) (U_h(x))^2\ ,
\lb{P2h2}
\ee
where $U_m(x)\,,\ m\ge 0\,$ are related to the Chebyshev polynomials of the second kind,
\be
U_m (2\,\cos t) = \frac{\sin m t}{\sin t}\qquad  \Rightarrow\quad U_m (2) = m\,,\quad U_2 (x) = x \ .
\lb{Um}
\ee
As it is easy to see, $U_m\,$ satisfy the recursion relation
\be
U_{m+1} (x) = x\, U_m (x) - U_{m-1} (x) \,,\quad m\ge 1\,,\quad U_0 (x) = 0\,,\quad U_1 (x) = 1
\lb{recurseUm}
\ee
so that all $U_m\,$ are {\em monic} polynomials and $\deg\, U_m = m-1\,.$
The equality $U_h (x)=\prod_{p=1}^{h-1} (x - \b_p )\,$ simply follows from here
since, by (\ref{Um}), $U_h(\b_p) = \frac{[hp]}{[p]}=0\,,\ 1\le p\le h-1\,.$
Eq. (\ref{P2h2}) implies that $C\,$ admits the following canonical (Jordan form) decomposition,
\be
C = \sum_{s=0}^{h} \b_s e_s + \sum_{p=1}^{h-1} w_p\,,\qquad
e_s e_{s'} = \d_{s s'} e_s\,,\qquad e_s w_p = \d_{s p} w_p\,,\qquad w_p w_{p'} = 0\,,
\lb{Cew}
\ee
in terms of $h+1\,$ central idempotents
$e_s \,,\ 0\le s\le h\,,\ \ \sum_{s=0}^h e_s = \id\,,\ $
and $h-1\,$ nilpotent central elements $w_p\,,\ 1\le p\le h-1\,,$ so that
\be
(C - \b_0 )\, e_0 = 0 = (C - \b_h )\, e_h\,,\qquad (C - \b_p )\, e_p = w_p\,,\qquad  (C - \b_p )\, w_p = 0\,.
\lb{wp}
\ee
The following standard consideration shows that the expansion (\ref{Cew}) is actually unique,
with $e_s\,,\ w_p\,$ (expressible as polynomials of degree $2h-1\,$ in $C$) satisfying (\ref{Cew}).
To this end we introduce the polynomials $Q^{(0)}(x)\,$ and $Q^{(h)}(x)\,$ (of degree $2h-1\,$) and
$Q^{(p)}(x)\,,\ 1\le p\le h-1\,$ (of degree $2h-2\,$), setting
\be
P_{2h} (x) = (x-\b_0 ) Q^{(0)}(x) = (x-\b_h ) Q^{(h)}(x) = (x-\b_p )^2 Q^{(p)}(x)\ ,\quad 1\le p\le h-1\ .
\lb{PQ}
\ee
It follows from (\ref{Cew}) that
\be
f(C) = \sum_{s=0}^{h} f(\b_s)\, e_s + \sum_{p=1}^{h-1} f'(\b_p)\, w_p\quad
\lb{fC}
\ee
for any (polynomial) function $f\,$ of $C\,.$
Using, further,
\be
Q^{(s)}(\b_r) = 0\quad{\rm for}\quad 0\le s\ne r\le h\ ,\qquad  (Q^{(p)})'(\b_r) = 0\quad{\rm for}\quad 1\le p\ne r\le h-1\ ,
\lb{Qsbr}
\ee
we find the relations
\be
Q^{(p)} (C) = Q^{(p)} (\b_p)\, e_p + (Q^{(p)})' (\b_p)\, w_p\,,\qquad (C-\b_p) Q^{(p)} (C) = Q^{(p)} (\b_p)\, w_p
\lb{ewC}
\ee
which one can solve for $w_p\,$ and $e_p\,.$

\medskip

The centre ${\cal Z}_q\,$ is not exhausted by the $2h$-dimensional space of polynomials of the Casimir operator.
The algebra $\bU\,$ admits a $\Z$-gradation such that
${\rm deg} (q^H) = 0\,,\ {\rm deg} (E) = 1\,,\ {\rm deg} (F) = -1\,$ and, due to (\ref{Uqres}),
only $2h-1\,$ of the homogeneous subspaces are nontrivial:
\be
\bU = \oplus_{\ell = 1-h}^{h-1} u^{(\ell )}_q\,,\qquad {\rm dim}\, u^{(\ell )}_q
= 2h (h- |\ell | )\,,
\lb{grad}
\ee
As $q^H x = q^{2\ell} x\, q^H\ \ \forall x\in u^{(\ell )}_q\,,$ it is clear that
${\cal Z}_q\,$ is the subalgebra of the $2 h^2$-dimensional algebra
\be
u^{(0)}_q\,= Span\, \{ E^r F^r q^{j H} \,,\ 0\le r\le h-1\,,\ 0\le j\le 2h-1 \}\,,
\lb{u0q}
\ee
singled out by the additional conditions $[E, z] = 0 = [F, z]\ \ \forall z\in {\cal Z}_q
\subset u^{(0)}_q\,.$

The characteristic equation for $q^H\in \bU\,,$
\be
q^{2 h H} = \id\qquad \Leftrightarrow\qquad \prod_{s=0}^{2h-1} (q^H - q^s) = 0\,,
\lb{Kcharact}
\ee
leads to its decomposition in terms of the idempotents $t_s\,$ projecting on the
eigenspaces corresponding to the eigenvalues $q^s\,,\ 0\le s\le 2h-1\,:$
\ba
&&q^H = \sum_{s\in{\Z} / {2h\,\Z}} q^s t_s\,,\quad t_s t_r = \d_{sr} t_s \qquad (\, \Rightarrow\  q^{jH} = \sum_{s\in{\Z} / {2h\,\Z}}q^{js} t_s\,,
\quad \sum_{s\in{\Z} / {2h\,\Z}} t_s = \id\, )\,,\nn\\
&&t_s = \frac{1}{2h}\,\sum_{j\in{\Z} / {2h\,\Z}}q^{-js} q^{jH}\,,\qquad
E\, t_s = t_{s+2\, mod\, 2h} E\,,\qquad F\, t_s = t_{s-2\, mod\, 2h} F\,.\qquad\qquad
\lb{K-idemp}
\ea
Introduce the projectors \cite{Ker, FGST1}
\be
\pi_p^+ = \sum_{m=0}^{p-1} t_{2m-p+1}\,,\qquad \pi_p^- = \sum_{m=p}^{h-1} t_{2m-p+1}\,,\qquad1\le p\le h-1
\lb{pipm}
\ee
(one has, in particular, $\pi_1^+ = t_0\,,\ \pi_{h-1}^- = t_h$)\,.
Note that $\pi_p^+\,$ projects exactly on
the eigenvectors of $q^H\,$ contained in $V^+_p\,,$ and $\pi_p^-\,$ -- on those in $V^-_{h-p}\,,$ since
$$\{ q^{2m -p + 1} \}_{m=p}^{h-1} = \{ - q^{2m - (h-p) + 1} \}_{m=0}^{h-p-1}\,.$$
The projectors $\pi_p^\pm\,$ themselves do not belong to ${\cal Z}_q\,,$ but one can check that the products
\be
w^\pm_p := \pi^\pm_p\, w_p\ ,\qquad 1\le p\le h-1
\lb{wp-pm}
\ee
do: $[E, w^\pm_p] = 0 = [F, w^\pm_p]\,.$
The relation $\sum_{m=0}^{h-1} q^{-2 jm} = h\, (\d_{j0} + \d_{jh})\,$ for $0\le j\le 2h-1\,$ implies
\be
\pi_p^+ + \pi_p^- = \sum_{m=0}^{h-1} t_{2m-p+1} = \frac{1}{2}\, (\id - (-1)^p q^{hH}) \,\in\, {\cal Z}_q\,,
\lb{pi+}
\ee
cf. (\ref{K-idemp}). Furthermore  (cf. Eqs. (\ref{qhHTh1}) and (\ref{wp})),
\ba
\lb{qhHonwp}
&&q^{hH} w_p = - T_h (\frac{1}{2}\, C)\, w_p = - T_h (\frac{1}{2}\, \b_p )\, w_p = - T_h ( \cos \frac{p \pi}{h} )\, w_p =
(-1)^{p-1} w_p\qquad\qquad\\ \nn\\
&&\qquad\Rightarrow\qquad w_p^+ + w_p^- = (\pi_p^+ + \pi_p^- )\, w_p = w_p\ .
\lb{w=sum}
\ea
Thus the $(3h-1)$-dimensional centre ${\cal Z}_q\,$ of $\bU\,$
is spanned by the $h+1\,$ idempotents
$e_0 , e_1, \dots , e_{h-1} , e_h \,,$ and the $2(h-1)\,$ nilpotent elements
$w^\pm_1 , \dots , w^\pm_{h-1}\,$ forming its {\em radical} (the largest nilpotent ideal):
\ba
&&e_r\, e_s = \d_{rs}\, e_r\ ,\qquad\quad 0\le r ,
s \le h\ ,\nn\\
&&e_r\, w_p^\pm = \d_{rp}\,w_p^\pm\ ,\qquad 0\le r \le h\,,\quad 1\le p \le h-1\ , \nn\\
&&w^\a_p w^\b_t = 0\ ,\qquad\qquad\, 1\le p\,, t \le h-1\ ,\quad \a,\b = \pm\ .\qquad\quad
\lb{ZbU}
\ea

The centre of $\bU\,$ is not represented faithfully in our Fock space
${\cal F}_q\,$ where, as it follows from Eq. (\ref{CF}), $C\,$
satisfies in fact the polynomial equation of degree $h+1\,$
\be
Q_{h+1}(C) = 0\,,\qquad Q_{h+1}(x):= \prod_{p=0}^h (x - \b_p ) = (x^2 - 4)\, U_h (x)\ .
\lb{1Qh+1}
\ee
(cf. (\ref{P2h2})). It is easily verified, by using (\ref{ewC}), (\ref{CF}) and (\ref{basis2}),
that the nilpotent elements $w_p\,$ (and hence, the whole radical)
annihilate any vector of the Fock space ${\cal F}_q\,.$
This means that the centre ${\cal Z}_q^F\,$ of the "radical free" algebra $U_q^F\,$
(the quotient of $\bU\,$ that is represented faithfully in our Fock space ${\cal F}_q\,$)
is spanned by the idempotents $\{ e_p \}_{p=0}^h\,$ alone. They can be now found from
\be
\left( \prod_{{s=0}\atop s\ne p}^h (\b_p-\b_s) \right)\, e_p = {\prod_{{s=0}\atop s\ne p}^h (C-\b_s)}\qquad\Rightarrow\qquad
(C - \b_p )\, e_p = 0\,,\qquad 0\le p\le h\,.
\lb{Qh+}
\ee

\medskip

\subsection{Drinfeld map of canonical $\bU\,$ characters}

A {\em balancing element} $g \in {\mathfrak A}\,$ is a group-like element,
$\,\Delta (g) = g\otimes g\,,$ satisfying
\be
S^2 (x) = g\, x\, g^{-1}\qquad\forall x\in {\mathfrak A}
\lb{balance}
\ee
(for a general Hopf algebra, its existence is not granted, and it may be not unique).
$\bU\,$ admits exactly two different balancing elements, $q^{H}\,$ and $q^{(h+1)H}\,$ (related
by multiplication with a central element); we shall choose in what follows $g = q^H\,.$

A {\em canonical ${\mathfrak A}$-character} (or {\em $q$-character}) $Ch_V^g\,$
is defined, for a given balancing element $g\,$
and any finite dimensional representation $\pi_V\,$ of ${\mathfrak A}\,,\ $  by
\be
Ch_V^g\, (x) := Tr_{\pi_V} (g^{-1} x)\qquad \forall x\in {\mathfrak A}\,.
\lb{canCh}
\ee
Any $q$-character satisfies the condition (\ref{Ch-Ad*inv}) (and hence, $Ch_V^g\in {\mathfrak C}{\mathfrak h}\,$):
\be
Ch_V^g\, (S^2(y) x) = Tr_{\pi_V} (g^{-1} S^2(y) x) = Tr_{\pi_V} (y g^{-1} x) =
Tr_{\pi_V} (g^{-1} x y) = Ch_V^g\, (xy)\ .
\lb{canch}
\ee

Note that both the Grothendieck ring (whose definition has been recalled in Section 2.3)
and the $q$-characters do not depend on the reducibility of the representations.
Moreover, the following property holds.

\medskip

\noindent
{\bf Proposition 3.2~} (\cite{Dr}) {\em The map of the Grothendieck ring of $\,{\mathfrak A}\,$ to the space
of $q$-characters given by $V \ \to\ Ch_V^g \subset {\mathfrak C}{\mathfrak h}\,$ is an algebra homomorphism.}

\medskip

\noindent
This means that, on top of the obvious relation $Ch_{V_1 + V_2}^g = Ch_{V_1}^g  + \, Ch_{V_2}^g\,,$ one has
\be
Ch_{V_1\otimes V_2}^g = Ch_{V_1}^g . \, Ch_{V_2}^g\,\qquad{\rm where}\quad
(\phi_1 . \phi_2 )(x) := (\phi_1 \otimes \phi_2 )\,\Delta (x)
\quad \forall\, \phi_1\,,\, \phi_2 \in {\mathfrak C}{\mathfrak h}\,.
\lb{V-Ch-homo}
\ee
The {\em proof} uses the identity
$\pi_{V_1\otimes V_2}  = (\pi_{V_1}\otimes\, \pi_{V_2})\, \Delta\,,$
the group-like property of the balancing element $g\,$ (\ref{balance}) implying
$\Delta (g^{-1} x) = (g^{-1}\otimes g^{-1} ) \Delta (x)\,,$
and the equality $Tr (A\otimes B) = Tr A \,\, Tr B\,.$

\medskip

The algebra of the $q$-characters of $\bU\,$
is a {\em proper subalgebra} of ${\mathfrak C}{\mathfrak h}\,$ in the sense that there are $\bU\,$
characters that are not traces of representations generated (by taking sums
and tensor products) from the set $\{ V^\epsilon_p \}\,$ (\ref{2h-IR})
of irreducible ones. Indeed, as it will become clear, the algebra of
$q$-characters of $\bU\,$ is isomorphic to the algebra generated by the
Casimir operator $C\,$ and hence has dimension $2h\,,$ while
${\mathfrak C}{\mathfrak h}\,,$ being isomorphic to the whole centre ${\cal Z}_q\,,$ is $(3h-1)$-dimensional.
Spanning the whole space of characters requires thus taking, in addition, into account some "pseudotraces"
(cf. \cite{FGST4, GT}) over (indecomposable) projective modules.

The existence of an $M$-matrix for $\bU\,$ allows to define
a map from the GR ${\mathfrak S}_{2h}\,$ to the centre ${\cal Z}_q\,$ of $\bU\,$ through
\be
D (V) := {\hat D} (Ch_V^g) \ \in\ {\cal Z}_q\,,
\lb{DPhi}
\ee
see (\ref{Dr-map}). Drinfeld's proof of (\ref{D-homom})
implies the following commutative algebra homomorphism ${\mathfrak S}_{2h}\ \rightarrow \ {\cal Z}_q\,:$
\be
D (V_1 . V_2) = {\hat D} (Ch_{V_1\otimes V_2}^g) = {\hat D} (Ch_{V_1}^g . \, Ch_{V_2}^g ) = D (V_1) . D(V_2)\,.
\lb{D-homom1}
\ee

\medskip

\noindent
{\bf Proposition 3.3~} (cf. \cite{FGST1}) {\em The Drinfeld images of the $\bU\,$ irreducible characters
\be
d^\epsilon_p := D (V^\epsilon_p) = ({\rm Tr}_{\pi_{V^\epsilon_p}} q^{-H}\otimes id ) \, {\cal M} \in {\cal Z}_q
\lb{Dr-Vp}
\ee
are given by}
\ba
&&d^+_p = \sum_{s=0}^{p-1} \sum_{\mu = 0}^s
\left[{{\mu+p-s-1}\atop{\mu}}\right] \left[{{s}\atop{\mu}}\right] \, \l^{2\mu} F^\mu E^\mu q^{(\mu +p-2s-1)(H+\mu+1)}\,,\nn\\
&&d^-_p = -\, q^{hH}\, d^+_p\,,\qquad 1\le p\le h \ .
\lb{DrVp2}
\ea

\medskip

\noindent
{\bf Proof~} To evaluate the traces in (\ref{Dr-Vp}), one should have in mind that
$Ch_V^g (x) \ne 0\,$ for $x\in u^{(0)}_q\,$ only, cf. (\ref{canCh}) and (\ref{u0q}), as well as
\be{\rm Tr}_{\pi_{V^\epsilon_p}} E^\mu F^\mu q^{jH} = \epsilon^{j+\mu} ([\mu]!)^2 \sum_{s=0}^{p-1} q^{j(2s-p+1)}
\left[{{\mu +p-s-1}\atop{\mu }}\right] \left[{{s}\atop{\mu }}\right]\ .
\lb{TrVa}
\ee
To prove (\ref{TrVa}), one uses (\ref{ErFr}), (\ref{CF}) and (\ref{Uqres-EFK}) to derive
\ba
&&E^\mu F^\mu q^{jH}\, | p, m{\cal i}^\epsilon = \frac{1}{\l^{2\mu}}\ q^{jH}\,\prod_{s=0}^{\mu -1}
(C - q^{H-2s-1} - q^{-H+2s+1}  )\,  | p, m{\cal i}^\epsilon = \nn\\
&&= \epsilon^{j+\mu}\, q^{j(2m-p+1)}\,
\prod_{s=0}^{\mu-1} \frac{ q^p+q^{-p}-q^{2(m-s)-p}-q^{p-2(m-s)} }{\l^{2}}\,| p, m{\cal i}^\epsilon =\nn\\
&&=\epsilon^{j+\mu}\, q^{j(2m-p+1)}\,\prod_{s=0}^{\mu -1}[p-m+s] [m-s]\,| p, m{\cal i}^\epsilon =\nn\\
&&= {\epsilon^{j+\mu}}([\mu ]!)^2 \, q^{j(2m-p+1)} \left[{{\mu +p-m-1}\atop{\mu }}\right] \left[{{m}\atop{\mu }}\right]
| p, m{\cal i}^\epsilon\,.
\lb{EFHa}
\ea
The Drinfeld maps $d^\epsilon_p= D (V^\epsilon_p)\,$ (\ref{Dr-Vp})
are thus given, in view of (\ref{Mmatr})
and (\ref{TrVa}), by
\ba
\lb{DrVp1}
&&d^\epsilon_p = \frac{1}{2h} \sum_{\mu=0}^{h-1} \sum_{m,n=0}^{2h-1} \frac{\l^{2\mu}\, q^{\mu }}{([\mu ]!)^2}\,
q^{mn + \mu (n-m)}\left( {\rm Tr}_{V^\epsilon_p} (E^\mu F^\mu q^{(m-1)H})\right)\, F^\mu E^\mu q^{nH} =\\
&&=\frac{1}{2h} \sum_{\mu=0}^{h-1} \sum_{m,n=0}^{2h-1} \epsilon^{\mu+m-1} q^{m(n-\mu)+\mu(n+1)} \l^{2\mu}
\sum_{s=0}^{p-1} q^{(m-1)(2s-p+1)}\left[{{\mu +p-s-1}\atop{\mu}}\right] \left[{{s}\atop{\mu}}\right] \, F^\mu E^\mu q^{nH}\,.
\nn
\ea
For $\epsilon = +1\,,$ taking the sum over $m\,$ makes the summation in $n\,$ automatic; on the other hand,
assuming $\epsilon = -1\ ( = q^h )\,$ is equivalent to multiplying the result for $\epsilon = +1\,$ by $-\, q^{hH}\,,$
arriving eventually at (\ref{DrVp2}). \eod

\medskip

We find, in particular,
\be
d^+_1 = \id \,,\qquad d^+_2 = C\,,\qquad d^-_1 = -\, q^{hH} = T_h (\frac{C}{2})\ .
\lb{Drinfeld12}
\ee
The result for $d^+_2\,$ could have been foreseen from (\ref{calcM}), since
\ba
&&d^+_2 \equiv ({\rm Tr}_{V^+_2}\otimes id)\,( ( q^{-H}\otimes \id )\, {\cal M}) = q^{\frac{3}{2}}\, {\rm Tr}\, (q^{-H^f} M) =
\lb{Tr1}\\
&&\nn\\
&&= {\rm Tr}\, \left\{ \left(\matrix{q^{-1}&0\cr0&q} \right)
\left(\matrix{q\l^2 FE + q^{-H} & -\l\, F q^H\cr - q \l\, E& q^H } \right) \right\} = \l^2 FE + q^{-H-1} + q^{H+1} = C\ ,
\nn
\ea
cf. (\ref{C}). Note that the alternative choice of the balancing element ($g=q^{H+1}\,,$ as in \cite{FGST1})
would lead to the opposite sign in (\ref{Tr1}), cf. (4.7) of \cite{FGST1}.

It turns out that the Drinfeld images of the canonical characters are insensitive
to the change ${\cal M}\ \leftrightarrow\ {\cal M}^{-1}\,.$ The fact that the expression
\ba
\lb{DrVp3}
&&({\rm Tr}_{\pi_{V^\epsilon_p}} q^{-H}\otimes id ) \, {\cal M}^{-1} =\\
&&=\frac{1}{2h} \sum_{\mu=0}^{h-1} \sum_{m,n=0}^{2h-1} \epsilon^{\mu+m-1} q^{m(\mu-n)-\mu(n+1)} \l^{2\mu}
\sum_{s=0}^{p-1} q^{(m-1)(p-2s-1)}\left[{{\mu + p-s-1}\atop{\mu}}\right] \left[{{s}\atop{\mu}}\right] \, E^\mu F^\mu q^{nH}
\nn
\ea
coincides with (\ref{DrVp1}) follows e.g. from the observation that it reproduces the values (\ref{Drinfeld12}) for
$d^\pm_1\,$ and $d^+_2\,$ (and the latter generate the algebra $D(V)\,$ (\ref{DPhi}), see the next subsection).
This is confirmed by the following trivial calculation:
\ba
&&q^{-\frac{3}{2}}\, {\rm Tr}\, (q^{-H^f} M^{-1}) = {\rm Tr}\, \left\{ \left(\matrix{q^{-1}&0\cr0&q} \right)
\left(\matrix{q^H & \l\,q^H F\cr q^{-1} \l\, E& q^{-1} \l^2 EF + q^{-H}} \right) \right\}=
\nn\\
&&\nn\\
&&= \l^2 EF + q^{H-1} + q^{1-H} = C \ (\,= d^+_2\, )\ .
\lb{Tr1-1}
\ea

\subsection{ The Grothendieck ring of $\bU\,$ and of $\tU\,$}

It has been shown in \cite{FGST1} that the multiplication rules for the GR ${\mathfrak S}_{2h}\,$
in terms of the irreducible representations of $\bU\,$ are
\ba
&&V_p^\epsilon\, .\, V_{p'}^{{\epsilon}'} =
\sum_{{s = |p - p'| + 1}\atop{s-p-p' = 1\, mod\, 2}}^{p+p'-1}\,
{\tilde{V}}^{\epsilon{\epsilon}'}_{s}\,,\qquad 1\le p\,,\,p' \le h\,,\ \ \epsilon\,,\,\epsilon' = \pm\,\,,\qquad\nn\\
&&{\tilde{V}}_s^\epsilon = \left\{
\begin{array}{ll}
V^\epsilon_s&{\rm for}\ 1\le s\le h\\
V^\epsilon_{2h-s} + 2\,V^{-\epsilon}_{s-h}&{\rm for}\ h+1\le s\le 2h-1
\end{array}
\right.\quad ,
\lb{GRres}
\ea
which can be equivalently encoded in the GR products of
$V_s^\pm\,,\ 1\le s\le h\,$ with $V^+_1\,$ and $V_2^+\,$ only; indeed,
the following subset of relations (\ref{GRres}) is sufficient
for the recursive reconstruction of the whole set:
\ba
&&V_1^+. V^\epsilon_1 = V^\epsilon_1\,,\qquad\qquad\qquad V_2^+. V^\epsilon_1 = V^\epsilon_2\,,\nn\\
&&V^\epsilon_p . V_2^+ = V^\epsilon_{p-1} + V^\epsilon_{p+1}\,,\qquad 2 \le p \le h-1\,,\qquad
V^\epsilon_h . V_2^+ = 2\, (V^\epsilon_{h-1} + V^{-\epsilon}_1 )\ .\qquad
\lb{GR-V2+}
\ea

The Fock space representation makes it natural to express the GR fusion
rules of both $\bU\,$ and $\tU\,$ in terms of the infinite number of representations ${{\cal V}_p}\,$ generated
by homogeneous polynomials of $a^1_\a\,$ of degree $p-1\,$ for $p=1, 2,\dots \,$ acting on the vacuum,
cf. (\ref{basis2}) albeit, except for the first $h\,$ values of $p\,,$ the latter are not irreducible.

\medskip
\noindent
{\bf Theorem 3.1~}

\noindent
{\em (a) The Grothendieck ring multiplication rules for ${\cal V}_p\,$ are of $su(2)$ type,
\be
{\cal V}_p \, . {\cal V}_{p'} = \sum_{{p^{\prime\prime} = |p - p'| + 1}\atop{p^{\prime\prime}-p-p'
= 1\, mod\, 2}}^{p+p'-1}\,
{\cal V}_{p^{\prime\prime}}\,,\qquad\qquad
p=1, 2,\dots \ .
\lb{n5}
\ee
Eqs. (\ref{GRres}) and (\ref{n5}) provide equivalent descriptions of the $\bU\,$ Grothendieck fusion algebra.}

\noindent
{\em (b) The Grothendieck ring of $\tU\,$ is equivalent to the "bosonic" subring of
${\mathfrak S}_{2h}\,$ containing an even number of negative parity irreducible modules
of $\bU\,$ (i.e., of type $V^-\,$).
This "parity rule" is respected by the decomposition (\ref{n5}).}

\medskip

\noindent
{\bf Remark 3.1~} The content of the right hand side of (\ref{n5}) can be already anticipated by
interpreting the results of \cite{PS, FK} about the tensor product expansion of two {\em irreducible} ${\cal V}_p\,$
(i.e., for $1\le p\le h\,$) "from the GR point of view", or by taking into account the well known fact that
a relation analogous to (\ref{n5}) holds (again for tensor products but this time without restriction on $p\,$)
for $q\,$ generic when all the representations ${\cal V}_p\,$ are irreducible;
in the GR context, one can expect it to remain true after specializing $q\,$ to a root of unity as well.

\medskip

\noindent
{\bf Proof~}

\noindent
(a) Since the subset of relations (\ref{n5}) for $p' =1, 2\,$ implies all the rest,
it is sufficient to prove only these using (\ref{GRpb}) and (\ref{GR-V2+}),
which is a straightforward calculation.
Conversely, to show that (\ref{GR-V2+}) (and hence, (\ref{GRres})) follow from (\ref{n5}),
one uses  (\ref{GRpb}) to express $V^-_p\,$ as
\be
V^-_p = \frac{1}{2}\, ({\cal V}_{h+p} - {\cal V}_{h-p})\,,\qquad 1\le p\le h\,.
\lb{V-calV}
\ee
To do this, one should consider (\ref{n5}) and (\ref{GRres}) -- defining, strictly speaking, a {\em semiring}
that can be extended, however, in a unique way to a true ring -- as relations in the Grothendieck fusion {\em algebra}.
For the present purpose it suffices to consider the latter over ${\Q}\,$ but,
having in mind the relation with ${\cal Z}_q\,,$ it is appropriate to extend it as an algebra over ${\C}\,.$

\noindent
(b) The first part of the proposition follows from the description, given in Section 2.3,
of the $\bU\,$ content of the $\tU\,$ irreducible modules, combined with the easily verifiable fact
that the additive subgroup of ${\mathfrak S}_{2h}\,$ characterized by an even number of IR of type $V^-\,$ is
also closed with respect to (Grothendieck) multiplication;
note that, in the case of $\tU\,,$ we have in mind the {\em true} ring
structure (as a "module over the integers") of ${\mathfrak S}_{2h}\,.$
The second assertion, which is in agreement with (\ref{V-calV}),
follows from the "bosonic nature" of ${\cal V}_p\,,$ see (\ref{GRpb}).
\eod

\medskip

It is clear that the relations contained in the infinite set (\ref{n5}) are not independent, and
the following construction (cf. \cite{FGST1}) illustrates this in a nice way.
Since the Drinfeld map $D\,$ relates, as an {\em isomorphism of associative commutative algebras}
\cite{Dr, Sch, FGST1}, ${\mathfrak S}_{2h}\,$ to the $2h$-dimensional
subalgebra of the centre generated by the Casimir operator (\ref{C}),
${\mathfrak S}_{2h} \ \stackrel{D}{\longrightarrow} \
{\mathfrak D}_{2h} \subset {\cal Z}_q\,,$
the algebra of the corresponding central elements provides
in the same time a description of the $\bU\,$ GR ${\mathfrak S}_{2h}\,.$
From (\ref{n5}) for $p' = 2\,$ and (\ref{Drinfeld12}) one derives
\be
D ({\cal V}_p ) = U_p (C)\,,\qquad p\in {\Z}_+\ ,
\lb{Dr-VP}
\ee
where $U_p (x)\,$ are the Chebyshev polynomials of the second kind (\ref{Um}) satisfying (\ref{recurseUm}).
Using (\ref{GRpb}) for $N=0\,$ and (\ref{V-calV}), one sees that the Drinfeld
images (\ref{Dr-Vp}) of the $\bU\,$ irreducible representations are given by
\be
d^+_p = U_p (C)\,,\qquad
d^-_p = \frac{1}{2}\, (U_{h+p} (C) - U_{h-p} (C) )\,,\qquad
1\le p\le h\ .
\lb{DR-gen}
\ee

\medskip

We end up this section with the proof of two important propositions announced in \cite{FGST1}. The following
characterization of the GR ${\mathfrak S}_{2h}\,$ provides an
important application of the expression (\ref{P2h2}) for $P_{2h}\,.$

\medskip

\noindent
{\bf  Proposition 3.4~}
{\em The Grothendieck ring ${\mathfrak S}_{2h}\,$ -- or, equivalently, its Drinfeld image ${\mathfrak D}_{2h} = D({\mathfrak S}_{2h})\,$ --
is isomorphic to the quotient of the algebra ${\mathbb C}\,${\rm [}$x${\rm ]} of polynomials of a single variable
with respect to the ideal, generated by
\be
P_{2h} (x) := U_{2h+1} (x) - U_{2h-1} (x) - 2 U_1 (x)
\equiv U_{2h+1} (x) - U_{2h-1} (x) - 2\,.
\lb{GR-Cheby}
\ee
The polynomial $P_{2h} (x)\,$ coincides with the one defined by (\ref{P2h=0}).
}
\medskip

\noindent
{\bf Proof~}
One can easily check indeed, see \cite{FGST1}, that with (\ref{DR-gen}) and (\ref{recurseUm})
one automatically reproduces all the relations (for the corresponding Drinfeld images) in (\ref{GR-V2+}),
except for the (last) one for $V^-_h .\, V_2^+\,$ which requires that the product
$$U_2(C)\, .\, \frac{1}{2}\, ( U_{2h}(C)- U_0(C)) \equiv
\frac{1}{2}\, ( U_{2h+1}(C)+ U_{2h-1}(C) )$$
should be equal to
$$( U_{2h-1}(C) - U_1(C) ) + 2 U_1(C) \equiv U_{2h-1}(C) + U_1(C)\ , $$
and hence, $P_{2h} (C) = 0\,.$ To demonstrate that the two
definitions (\ref{GR-Cheby}) and (\ref{P2h=0}) of $P_{2h} (x)\,$ coincide, one uses
(\ref{Um}) and (\ref{Cheby1}) to derive
\be
U_{m+1}(2 \cos t) - U_{m-1}(2\cos t) = 2\, T_m (\cos t)\ ,
\lb{psi-cos}
\ee
and then applies (\ref{P2hT2h}).\eod

\medskip

The decomposition (\ref{Cew}) shows that the
subalgebra ${\mathfrak D}_{2h} \subset {\cal Z}_q\,$
is generated by $\{ e_s \}_{s=0}^h\,$ and
$\{ w_p \}_{p=1}^{h-1}\,.$ Its radical,
hence,  is of dimension $(h-1)\,,$
being spanned by $\{ w_p \}_{p=1}^{h-1}\,.$
The {\em annihilator} ${\mathfrak K}_{h+1}\,$ of the radical of ${\mathfrak D}_{2h}\,$
is $(h+1)$-dimensional, including the elements of the radical itself,
and also $e_0\,$ and $e_h\,,$ see (\ref{ZbU}).
Let  ${\cal K}_{h+1}\,$ be the ideal of the GR ${\mathfrak S}_{2h}\,$
spanned by
\be
K_r := V^+_r + V^-_{h-r}\ ,\qquad 0\le r\le h\ ,
\lb{Verma-ideal}
\ee
($K_0 \equiv V^-_h \,$ and $K_h \equiv V^+_h\,$ since $V^{\pm} = \{ 0 \}\,,$ cf. (\ref{epN})).
${\cal K}_{h+1}\,$ is the "Verma-module ideal" of \cite{FGST1}; it is easy to see,
using (\ref{GR-V2+}), that the Grothendieck products with $V^+_2\,$
of the set of the $h$-dimensional spaces
(\ref{Verma-ideal}) are expressed again as sums of the latter, so that indeed
${\mathfrak S}_{2h}\, .\, {\cal K}_{h+1}\subset {\cal K}_{h+1}$.
Using  (\ref{fC}) and (\ref{DR-gen}), one can prove the following

\medskip

\noindent
{\bf Proposition 3.5~}
{\em The Drinfeld map $D({{\cal K}}_{h+1})\,$ of the Verma-module ideal
coincides with the annihilator ${\mathfrak K}_{h+1}\,$ of the radical of $\ {\mathfrak D}_{2h}\,.$
The quotient ${\mathfrak D}_{2h} / {\mathfrak K}_{h+1}\,$ is isomorphic to the
fusion ring of the unitary $\widehat{su}(2)_{h-2}\,$ WZNW model.}

\medskip

\noindent
{\bf Proof~} Applying (\ref{DR-gen}), one gets for the Drinfeld images $\kappa_r := D(K_r)\,$
\ba
&&\kappa_p  = d^+_p + d^-_{h-p} = \frac{1}{2}\, (U_p (C) + U_{2h-p} (C) )\,,\qquad 1\le p\le h-1\,,\nn\\
&&\kappa_h = d^+_h = U_h (C)\,,\qquad \kappa_0 = d^-_h  = \frac{1}{2}\,U_{2h}(C)
\lb{D-kappa1}
\ea
or, equivalently,
\be
\lb{D-kappa2}
\kappa_r = \frac{1}{2}\,\sum_{s=0}^h (U_{2h-r} (\b_s) + U_r (\b_s))\, e_s
+ \frac{1}{2}\,\sum_{t=1}^{h-1} ({U}^{~'}_{2h-r} (\b_t) + {U}^{~'}_r (\b_t))\, w_t \,,\quad 0\le r\le h
\ee
(cf. (\ref{fC})). Since, for any $r\,,\ U_r (\b_s) = U_r (2\cos\frac{s\pi}{h}) = \frac{[r s]}{[s]}\,,$
it follows that
\be
U_h ({\b}_s) = 0 =  U_{2h} ({\b}_s)\,,\qquad 1\le s\le h-1
\lb{U0p}
\ee
and
\be
U_p (\b_s) + U_{2h-p} (\b_s) = \frac{[p\, s] + [(2h-p)s]}{[s]} = 0\,,\qquad 1\le p\,, s\le h-1\ .
\lb{Up-p}
\ee
This means that all the coefficients of $\{ e_s \}_{s=1}^{h-1}\,$ in the right hand sides of
Eqs. (\ref{D-kappa2}) are zero, so that any element of $D({\cal K}_{h+1})\,$
annihilates the radical of ${\mathfrak D}_{2h}\,.$
On the other hand, the coefficients of $e_0\,$ and $e_h\,$ in (\ref{D-kappa2}) are simply given by
\be
\frac{1}{2}\, (U_{2h-r} (2) + U_r (2)) = h\ ,\qquad\frac{1}{2}\, (U_{2h-r} (-2) + U_r (-2)) = (-1)^{r-1} h\ ,
\lb{coeffeh0}
\ee
respectively. From (\ref{Um}) we get
${U}^{~'}_r (\b_t) = \frac{r\, \b_{r t} - \b_t \frac{[r t]}{[t]}}{(q^t-q^{-t})^2 }\ ,$
so that the Drinfeld images (\ref{D-kappa1}) are given by
\be
\kappa_r  = h  \left( e_0 + (-1)^{r-1} e_h + \l^{-2} \sum_{t=1}^{h-1} \frac{\b_{r t}}{[t]^2}\, w_t \right)\,,\qquad 0\le r\le h\ .
\lb{D-kappa3}
\ee
The $(h+1)\times (h+1)\,$ symmetric matrix
\be
A(h) = ( A_{rs} )\,,\qquad A_{rs}= \cos \frac{rs \pi}{h}\,,\quad r,s = 0,1,\dots , h
\lb{matrAh}
\ee
is invertible, as its determinant\footnote{The authors thank Alexander Hadjiivanov for suggesting
formula (\ref{det A}). It has been then numerically verified up to large values of $h\,.$}
\be
\det A(h) = (-1)^{\frac{h(h+1)}{2} }\,  2^{\frac{3-h}{2}}\,  h^{\frac{h+1}{2}}
= 4\, \left( \frac{(-1)^h h}{2} \right)^{\frac{h+1}{2}}
\lb{det A}
\ee
does not vanish. Thus the set $\{ \kappa_r \}_{r=0}^h\,$
provides another basis of the annihilator ${\mathfrak K}_{h+1}\,$ of the radical of ${\mathfrak D}_{2h}\,,$
i.e., $D({\cal K}_{h+1}) = {\mathfrak K}_{h+1}\,.$
The $(h-1)$-dimensional quotient ${\mathfrak D}_{2h} / {\mathfrak K}_{h+1}\,$
is spanned therefore by the equivalence classes of elements
$e_p + x\,,\ 1\le p\le h-1\,,\ x\in {\mathfrak K}_{h+1}\,$
and the {\em canonical images} $D_p = \sum_{s=1}^{h-1} \frac{[ p\, s ]}{[s]}\,,\ 1\le p\le h-1\,$
of $D (V_p^+ )\,$ in the quotient form a basis in ${\mathfrak D}_{2h} / {\mathfrak K}_{h+1}\,.$
Due to the well-known property of the quantum brackets, $D_p \,$ reproduce the
fusion rules
\be
D_p\,  D_{p'} = \sum_{{p'' = |p-p'|+1}\atop{p'' -p-p'=1\, mod\, 2}}^{h-1-|h-p-p'|} \, D_{p''}\,,
\qquad 1\le p,\, p' \le h-1
\lb{fusion-su2}
\ee
for primary fields of isospins $0\le I,\,I' \le \frac{h}{2} -1\,$
in the unitary $\widehat{su}(2)_{h-2}\,$ WZNW model for $ p=2I+1\,,\ p' = 2I'+1\,.$\eod

\medskip

\section{Duality between $\tU\,$ and braid group representations}

\setcounter{equation}{0}
\renewcommand\theequation{\thesection.\arabic{equation}}

\medskip

\subsection{Statement of the problem. Definition of the braid group module ${\cal S}_4(p)$}

It follows from the very definition of the $R$-matrix as an intertwiner between the, say $U_q\,,$ coproduct
and its opposite, and of the braid operator, $\hat R = R P\,,$ where $P\,$ stands for permutation, that
the braid group ${\cal B}_n\,$ realized in the $n$-fold tensor product of a $\tU\,$ module ${\cal V}_p\,,$
belongs to the commutant of the $\tU\,$ action. This provides a correspondence between the representations of $\tU\,$
and ${\cal B}_n\,$ which can be viewed as a deformation of the well known Schur-Weyl duality
between the representations of $U(k)\,$ (or $GL(k, {\mathbb C})\,$) and the permutation group ${\cal S}_n\,,$ both acting in the
$n$-fold tensor power of ${\mathbb C}^k\,.$

The knowledge of the $p$-dimensional realization of ${\cal B}_4\,$ in the space of $su(2)\,$
current algebra blocks allows to establish another type of duality relation between
QUEA and braid group representations. Such a relation has been studied in
the case of (unitary) irreducible representations of $\tU\,,$ corresponding
to integrable ${\widehat{su}}(2)_{h-2}\,$ modules. The set of associated $\tU$-covariant chiral primary fields, however,
is not closed under fusion. Thus, we have to study the above relation for {\em indecomposable} current algebra and $\tU\,$
modules as well. We do this in Section 4.2 for the group ${\cal B}_4\,$ acting on the space ${\cal S}_4(p)\,$ of $4$-point
blocks which we proceed to introduce.

Let $F(p)\,$ be a $p$-dimensional (unitary) irreducible $SU(2)\,$ module characterized by its isospin $I_p\,$ or weight
$2I_p = p-1\,,$ and $J(p) = Inv\, (\, F(p)^{\otimes 4}\, )\,$ -- the $p$-dimensional space
of invariant tensors in the $4$-fold tensor product of $F(p)\,.$ The space ${\cal S}_4(p)\,$
is defined as the space of tensor valued functions $w = w(z_1\,,\,\dots\,,\, z_4)\in J(p)\,$ which

\vspace{1mm}

\noindent
(i) satisfy the KZ equation
\be
\left(\, h\,\frac{\partial}{\partial z_a} - \sum_{{b=1}\atop{b\ne a}}^4 \frac{\Omega_{ab}}{z_{ab}}\,\right)\,
w(z_1\,,\,\dots\,,\, z_4) = 0\,,\qquad z_{ab}=z_a-z_b\,,\qquad \Omega_{ab}=\Omega_{ba}\ ,
\lb{KZE}
\ee
where $\Omega_{ab} = 2 \overrightarrow{I_a}\, .\, \overrightarrow{I_b}\,$ are the polarized Casimir operators
acting nontrivially only on the tensor product $F_a(p) \otimes F_b(p)\,;$

\vspace{1mm}

\noindent
(ii) are M\"obius invariant with respect to the tensor product of positive energy representations of $SU(1,1)\,$ of
conformal weight (minimal energy)
\be
\Delta_p = \frac{I_p (I_p+1)}{h} = \frac{p^2-1}{4h}\ .
\lb{DIp}
\ee

\vspace{1mm}

It follows from these assumptions that the function $w\,$ is, in general, multivalued analytic with possible singularities
(branch points) at coinciding arguments. We define the principal branch of $w\,$ as the solution in a complex neighbourhood
of the real open set $z_1 > z_2 > z_3 > z_4 > 0\,$ given by a convergent Taylor expansion in the variables $\frac{z_a+1}{z_a}\,,\
a=1,2,3\,.$ This allows to define, using analytic continuation along appropriate homotopy classes of paths, a ($p$-dependent)
monodromy representation of the braid group ${\cal B}_4\,$ of four strands. We shall demonstrate in Section 4.2 that for
any $N\ge 1\,$ and $1\le p \le h-1\,$ the ${\cal B}_4\,$ module ${\cal S}_4(Nh+p)\,$ admits an $N (h-p)$-dimensional
braid invariant submodule. Moreover, we shall establish a precise duality between the (indecomposable) structure of
${\cal S}_4(p)\,, \ p=1, 2,\dots\, $ and that of the homogeneous subspaces
${\cal V}_p\,$ of the zero modes Fock space ${\cal F}_q\,.$

\medskip

\subsection{Indecomposable structure of the ${\cal B}_4\,$ module ${\cal S}_4(p)\,$}

We shall use the methods and results of \cite{MST} and \cite{STH} to write down explicitly the ${\cal B}_4\,$ action
on ${\cal S}_4(p)\,.$ To begin with, we shall view each $F_a(p)\,,\ a=1,2,3,4\,$ as a space of polynomials of degree
$(2I = )\, p-1\,$ in a variable $\zeta_a\,.$ Then the $SU(2)$-invariant $4$-point blocks $w\,$ appear as homogeneous
polynomials of degree $2(p-1)\,$ in the differences $\zeta_{ab} = \zeta_a-\zeta_b\,.$ We can express $w\,$ in terms
of an amplitude $f\,$ that depends on two invariant cross ratios, writing
\be
w (\zeta_1, z_1 ; \dots ; \zeta_4 , z_4 )= D_p\, ( \zeta_{ab}\,,\, z_{ab} ) \, f(\xi, \i)\,,\qquad
D_p = \left(\frac{z_{13}z_{24}}{z_{12}z_{34}z_{14}z_{23}} \right)^{2\Delta_p} (\zeta_{13}\zeta_{24})^{p-1}\ ,
\lb{wDpf}
\ee
where $z_{ab} = z_a-z_b\,,$
\be
\xi = \frac{\zeta_{12}\zeta_{34}}{\zeta_{13}\zeta_{24}}\,,\qquad \i = \frac{z_{12}z_{34}}{z_{13}z_{24}}\ ,
\lb{xi}
\ee
and $f\,$ is a polynomial in $\xi\,$ of degree not exceeding $p-1\,.$
The Casimir operators are then transformed into differential operators in $\xi\,$ and the KZ equation (\ref{KZE})
assumes the form
\be
\left(\, h\,\frac{\partial}{\partial \i} - \frac{C_{12}}{\i} + \frac{C_{23}}{1-\i}\, \right)  f = 0\,,\qquad
C_{ab} = ( \overrightarrow{I_a}+ \overrightarrow{I_b} )^2 = \Omega_{ab}+\frac{p^2-1}{2}
\lb{KZf}
\ee
or, explicitly,
\ba
&&C_{12} = (p-1)( p-(p-1)\,\xi) - (2(p-1)(1-\xi)+\xi )\,\xi\frac{\partial}{\partial \xi} +
\xi^2(1-\xi)\frac{\partial^2}{\partial \xi^2}\ ,\nn\\
&&C_{23} = (p-1) ( p-(p-1)(1-\xi) ) + ( 2(p-1)\xi+1-\xi )(1-\xi)\frac{\partial}{\partial \xi} +
\xi (1-\xi)^2\frac{\partial^2}{\partial \xi^2}\ .
\qquad\quad
\lb{C12C23}
\ea

A {\em regular basis} of the $p\,$ linearly independent solutions
$\{\, f^{(p)}_\mu = f_\mu (\xi , \i)\,,\quad \mu=0,1,\dots ,\\ p-1 \}\,$
of Eqs. (\ref{KZf}), (\ref{C12C23}) has been constructed in \cite{STH} in terms of appropriate multiple
contour integrals. We shall only use here the {\em well defined}\footnote{By contrast, the commonly used "$s$-basis"
that pretends to diagonalize the braid matrix $B_1\,$ actually does not exist, yielding
"singular braid matrices" for $p\ge h\,.$} explicit braid group action on $f_\mu\,$ (and $w_\mu = D_p\, f_\mu\,$).
If $B_i\,$ corresponds to a rotation of the pair of world sheet variables
$(z_i\,,\, z_{i+1} )\,$ at an angle $\pi\,$ in the positive direction and a simultaneous exchange
$\zeta_i\ {\rightleftarrows}\  \zeta_{i+1}\,,$ we have
\ba
&&B_1 \ ( = B_3)\, :\, f_\mu (\xi, \i)\ \rightarrow\ (1-\xi)^{p-1}(1-\i)^{4\Delta_p}
f_\mu(\frac{\xi}{\xi-1},\frac{e^{-i\pi}\i}{1-\i}) = f_\l(\xi,\i)\,{B_1}^\l_{~\mu}\ ,\nn\\
&&B_2\, :\, f_\mu (\xi, \i)\ \rightarrow\ \xi^{p-1}\i^{4\Delta_p}
f_\mu(\frac{1}{\xi},\frac{1}{\i}) = f_\l(\xi,\i)\,{B_2}^\l_{~\mu}\ ,
\lb{B1B3B2}
\ea
where $({B_i}^\l_{~\mu})\,,\ i=1,2\,$ are (lower and upper, respectively) triangular $p\times p\,$ matrices:
\ba
&&{B_1}^\l_{~\mu} =  (-1)^{p-\l-1} q^{\l (\mu +1)-\frac{p^2-1}{2}}
\left[{\l\atop\mu}\right]\ ,\nn\\
&&{B_2}^\l_{~\mu} = (-1)^\l q^{(p-\l-1) (p-\mu )-\frac{p^2-1}{2}}
\left[{p-\l-1\atop p-\mu-1}\right]\ ,\qquad \l\,,\ \mu = 0,1,\dots , p-1\ .\qquad
\lb{B1B2}
\ea

We shall be only interested in what follows in the representations of ${\cal B}_4\,$ (equivalent to those) explicitly
given by (\ref{B1B2}). They can be partly characterized by the condition on the eigenvalues of the generators
\be
({B_i}^\l_{~\l} )^{4h} = 1\ ,\qquad i = 1,2,3\,,\qquad \l = 0, 1, \dots , p-1
\lb{Bev}
\ee
on top of the general requirements
\ba
&&B_i B_{i+1} B_i = B_{i+1} B_i B_{i+1}\,,\quad i=1, 2\ ,\qquad\quad B_1 = B_3\ , \nn\\
&&B_1 B_2 B_3^2 B_2 B_1 \equiv (B_1 B_2 B_1)^2 = (B_2 B_1 B_2)^2 = (-1)^{p-1} q^{1-p^2} \id\ .
\lb{BG}
\ea

\medskip

\noindent
{\bf Theorem 4.1~}
{\em The $p$-dimensional ${\cal B}_4\,$ modules ${\cal S}_4(p)\,$ have a structure
dual to that of the $\tU\,$ modules ${\cal V}_p\,$ in the following sense.

\noindent
(a) ${\cal S}_4(p)\,$ are irreducible for $1\le p\le h-1\,$ and for $p=Nh\,,\ N\ge 1\,.$

\noindent
(b) For $N\ge 1\,,\ 1\le p \le h-1\,,$ ${\cal S}_4(Nh+p)\,$ is indecomposable, with structure "dual" to that
displayed by the exact sequence (\ref{shex-eqN}):
\be
0\ \ \rightarrow\ S(N,h-p) \ \ \rightarrow\ \ {\cal S}_4(Nh+p)\ \
\rightarrow\ \  {\tilde S}(N+1, p)\ \ \rightarrow\ \ 0\ ,
\lb{shex-eqS}
\ee
where $S (N,h-p)\,$ is the $N (h-p)$-dimensional invariant subspace of ${\cal S}_4(Nh+p)\,$ spanned by the vectors
\be
S(N,h-p) = Span \, \{\,f_\mu^{(Nh+p)}\,,\ \mu = nh+p\, , \dots , (n+1)h-1\, \}_{n=0}^{N-1}\ ,
\lb{Sh-p}
\ee
which carries an IR of ${\cal B}_4\,;$ the $(N+1)p$-dimensional quotient space ${\tilde S}(N+1, p)\,$
also carries an IR of the braid group.
}

\medskip

\noindent
{\bf Proof~} The fact that the subspace $S(N, h-p)\,$ (\ref{Sh-p}) is ${\cal B}_4\,$ invariant
follows from the observation that the ($(Nh+p)$-dimensional) matrices (\ref{B1B2}) satisfy
\ba
&&{B_1}^{mh+\a}_{~nh+\b}\ \sim \ \left[{{mh+\a}\atop{nh+\b}}\right] \ \sim\
\left[{\a\atop \b}\right] \left({m\atop{n}}\right) = 0\ ,\lb{B=0}\\
&&{B_2}^{mh+\a}_{~nh+\b}\ \sim \ \left[{{(N-m)h+p-\a-1}\atop{(N-n-1)h+h+p-\b-1}}\right] \ \sim\
\left[{p-\a-1\atop{h+p-\b-1}}\right] \left({N-m\atop{N-n-1}}\right)  = 0\nn
\ea
for $m=0\, , \dots , N , \ \ 0\le \a\le p-1\,$ and
$n=0\, , \dots , N-1\,, \ \ p\le \b \le h-1\,,$ cf. (\ref{q-bin1}) (since $\a<\b\,$ and $p-\a-1<h+p-\b-1\,$).
An inspection of the same expressions (\ref{B1B2}) allows to conclude that the space
$S(N, h-p)\,$ has no ${\cal B}_4\,$ invariant complement in ${\cal S}_4 (Nh+p)\,$ which is, thus, indeed indecomposable.
It is also straightforward to verify that the quotient space
\be
{\cal S}_4(Nh+p) / S(N, h-p)\ \simeq\ {\tilde S}(N+1, p)
\lb{factS}
\ee
carries an IR of ${\cal B}_4\,.$ \eod

\medskip

We thus see that the indecomposable representations ${\cal V}_{Nh+p}\,$ (of $\tU\,$) and ${\cal S}_4(Nh+p)\,$
(of ${\cal B}_4\,$) contain the same number (two) of irreducible components
(of the same dimensions), but the arrows of the exact sequences
(\ref{shex-eqN}) and (\ref{shex-eqS}) are reversed. This sums up the meaning of duality
for indecomposable representations.

For $p=h-1\,,$ the ${\cal B}_4\,$ invariance and irreducibility of the space $S(N,1)\,$ spanned by
$\{\, f_\mu^{(mh-1)}\, \}_{m=1}^{N}\,$ (cf. (\ref{Sh-p})) has been displayed by A. Nichols in \cite{N1}.

\medskip

\noindent
{\bf Remark 4.1~} We note that the difference of conformal dimensions
\be
\Delta_{2Nh+p} - \Delta_p = N(Nh+p)\,, \qquad 1\le p\le h-1\,,\quad N=1, 2, \dots
\lb{D-int}
\ee
is a (positive) integer, which explains the similarity of the corresponding braid group representations
${\cal S}_4 (p)\,$ and ${\cal S}_4 (2Nh+p)\,.$

\medskip

\noindent
{\bf Remark 4.2~} There is a unique $1$-dimensional subspace $S(1,1) \in {\cal S}_4 (2h-1)\,$
among the ${\cal B}_4\,$ invariant subspaces displayed in Theorem 4.1 corresponding to a
(non-unitary) local field of isospin and conformal dimension $h-1:$
\be
\Delta_{2h-1} = \frac{(2h-1)^2 -1}{4h} = h-1 = \frac{I(I+1)}{h}\ \ {\rm for}\ \ I=h-1\,\ .
\lb{DI}
\ee
It has rational correlation functions (in particular, the $4$-point amplitude $f^{(2h-1)}_{h-1}
(\xi , \i )\,$ is a polynomial \cite{HP}) and gives rise to a local extension of the $\widehat{su}(2)_{h-2}\,$ current algebra
whose superselection sectors involve direct sums of the type
\be
{\cal S}_4(p) \oplus S(1,p)\quad{\rm for}\quad 1\le p\le h-1\ ,
\lb{dirsumS}
\ee
where $S(1,p)\,$ is the $p$-dimensional invariant subspace of ${\cal S}_4(2h-p)\,.$

\medskip

\subsection{Concluding remarks}

It has been argued in \cite{FGST1, FGST2} that a {\em Kazhdan-Lusztig} type correspondence
holds between the Grothendieck fusion rings of the logarithmic $c_{1 p}\,$ minimal model
and of the restricted QUEA $\bU\,.$ Theorem 4.1, proven in the preceding subsection,
indicates the presence of a more precise duality relation between monodromy representations of the braid group acting
on solutions of the KZ equation and the Fock space representations ${\cal V}_p\,$ of $\tU\,,$ involving arrow reversal
in the exact sequences describing the corresponding indecomposable structures. This result should motivate further study
of chiral current algebra models beyond the unitarity limit that may reveal the CFT counterpart of more structures of
the $\tU\,$ tensor category (such as the Drinfeld map) studied in Section 3.

\medskip

\section*{Acknowledgments}

The authors thank a referee for his stimulating criticism.
L.H. thanks Dr. L. Georgiev for computer assistance.
Parts of this work has been done during visits of L.H. at the International School for Advanced Study
(SISSA/ISAS), supported in part by the Central European Initiative (CEI),
and at Istituto Nazionale di Fisica Nucleare (INFN), Sezione di Trieste, of I.T. at
the Abdus Salam International Centre for Theoretical Physics (ICTP) in Trieste and at CERN, and during an INFN supported visit
of P.F. at the Institute for Nuclear Research and Nuclear Energy (INRNE) in Sofia. The authors thank these institutions
for hospitality and support. The work of L.H. and I.T. has been supported in part by the
Bulgarian National Foundation for Scientific Research (contract Ph-1406) and
by the European RTN EUCLID (contract HPRN-CT-2002-00325) and Forces-Universe (contract MRTN-CT-2004-005104).
P.F. acknowledges the support of the Italian Ministry of University and Research (MIUR).

\end{document}